\documentclass[nofootinbib,aps,preprint,prc]{revtex4}
\usepackage[utf8]{inputenc}

\usepackage{amsmath}
\usepackage{mathrsfs}
\usepackage{amssymb}
\usepackage{bbm}
\usepackage{graphicx}
\usepackage{graphics}
\usepackage[small,compact]{titlesec}
\usepackage{xspace}
\usepackage{colordvi}
\usepackage{color}
\usepackage{units}
\usepackage{stackrel}
\usepackage{hyperref}
\usepackage{slashed}

\usepackage{ulem}

\renewcommand{\emph}[1]{\textit{#1}}

\newcommand{\vect}[1]{\vec{\mathbf{#1}}}

\newcommand{\EFT}{$\mathrm{EFT}(\slashed{\pi})$\xspace}

\newcommand{\comment}[1]{}
\newcommand{\srutTypeOne}[1]{\vrule width0pt height0pt depth #1\relax}
\newcommand{\jjvHe}{{}^3\mathrm{He}}
\newcommand{\jjvH}{{}^3\mathrm{H}}
\newcommand{\nnlo}{\mathrm{NNLO}}

\newcommand{\G}{\mathcal{G}}
\newcommand{\Gb}{\boldsymbol{\mathcal{G}}}
\newcommand{\Dwd}{\widetilde{D}}

\makeatletter

\newcommand{\Rmnum}[1]{\expandafter\@slowromancap\romannumeral #1@}
\newcommand{\vast}{\bBigg@{4}}
\newcommand{\Vast}{\bBigg@{5}}
\makeatother

\begin{document}

\title{Charge and Magnetic Properties of Three-Nucleon Systems in Pionless Effective Field Theory}

\author{Jared Vanasse}
\email{vanasse@ohio.edu}
\affiliation{Institute of Nuclear and Particle Physics and Department of Physics and Astronomy Ohio University, Athens OH 45701, USA
}

\date{\today}

\begin{abstract}
A method to calculate the form factor for an external current with non-derivative coupling for the three-body system in an effective field theory (EFT) of short-range interactions is shown.  Using this method the point charge radius of $\jjvHe$ is calculated to next-to-next-to-leading order ($\nnlo$) in pionless EFT (\EFT), and the magnetic moment and magnetic radius of $\jjvH$ and $\jjvHe$ are calculated to next-to-leading order (NLO).  For the $\jjvHe$ charge and magnetic form factors Coulomb interactions are ignored.  The $\jjvHe$ point charge radius is given by 1.74(4)~fm at $\nnlo$.  This agrees well with the experimental $\jjvHe$ point charge radius of 1.7753(54)~fm~\cite{Angeli201369}.  The $\jjvH$ ($\jjvHe$) magnetic moment in units of nuclear magnetons is found to be 2.92(35) (-2.08(25)) at NLO in agreement with the experimental value of 2.979 (-2.127).  For $\jjvH$ ($\jjvHe$) the NLO magnetic radius is 1.78(11)~fm  (1.85(11)~fm) which agrees  with the experimental value of 1.840(182)~fm (1.965(154)~fm)~\cite{Sick:2001rh}.  The fitting of the low-energy constant $L_{1}$ of the isovector two-body magnetic current and the consequences of Wigner-SU(4) symmetry for the three-nucleon magnetic moments are also discussed.
\end{abstract}


\maketitle

\section{Introduction}

When systems are probed at length scales much larger than the scale of their underlying interaction $r$ then those interactions can be expanded in a series of contact interactions known as short range effective field theory (srEFT).  Systems with short range interactions (\emph{i.e.} cold atom systems, halo nuclei, and low energy few-nucleon systems) exhibit such behavior at low energies.  The applicability of srEFT to such a broad class of systems is known as universality~\cite{Braaten:2004rn}.  Importantly, srEFT possesses a power counting that allows for systematically improvable calculations with error estimates.  The power counting is in powers of $(Q/\Lambda)^{n}$, where $Q$ is the typical momentum scale of particles in the system, $\Lambda\sim 1/r$ is the breakdown scale of srEFT, and using naive dimensional analysis~\cite{vanKolck:1998bw} low energy constants (LECs) in the theory are assumed to scale dimensionally in powers of $\Lambda$.  However, for physical systems of interest it is observed that the scattering length $a$ scales unnaturally ($r<a\sim 1/Q$).  This leads to interactions in $a$ being treated non-perturbatively at leading order (LO) and the creation of relatively shallow two-body bound states~\cite{Kaplan:1998tg,Kaplan:1998we}.  Higher order range corrections are then added perturbatively in a series of $r/a\sim Q/\Lambda$.

srEFT has been used successfully in the description of low-energy few-nucleon systems through the use of pionless EFT (\EFT), characterized by the breakdown scale $\Lambda_{\not{\pi}}\sim m_{\pi}$ and valid for energies $E<m_{\pi}^{2}/M_{N}$. \EFT has been used in the two-body sector to calculate nucleon-nucleon ($N\!N$) scattering~\cite{Chen:1999tn,Kong:1998sx,Kong:1999sf,Ando:2007fh}, neutron-proton ($np$) capture~\cite{Chen:1999tn,Chen:1999bg,Ando:2004mm} to ($\lesssim1\%$)~\cite{Rupak:1999rk}, deuteron electromagnetic properties~\cite{Chen:1999bg,Ando:2004mm}, proton-proton fusion~\cite{Kong:2000px,Ando:2008va,Chen:2012hm}, and neutrino-deuteron scattering~\cite{Butler:2000zp}.  In the three-body sector it has been used to calculate neutron-deuteron ($nd$) scattering ~\cite{Bedaque:1998mb,Bedaque:1999ve,Gabbiani:1999yv,Bedaque:2002yg,Griesshammer:2004pe,Vanasse:2013sda,Margaryan:2015rzg}, proton-deuteron ($pd$) scattering~\cite{Rupak:2001ci,Konig:2011yq,Konig:2013cia,Vanasse:2014kxa,Konig:2014ufa,Konig:2016iny}, $\jjvH$ and $\jjvHe$ binding energies~\cite{Bedaque:1999ve,Ando:2010wq,Konig:2011yq,Konig:2015aka}, three-nucleon electromagnetic~\cite{Platter:2005sj,Kirscher:2017fqc} and weak properties~\cite{De-Leon:2016wyu}, and $nd$ capture~\cite{Sadeghi:2006fc,Arani:2014qsa}.

Techniques to calculate $nd$ scattering  strictly perturbatively were introduced in Ref.~\cite{Vanasse:2013sda}.  Ref.~\cite{Vanasse:2015fph} then extended this method to the calculation of perturbative corrections to three-body bound states.  Using these methods, Ref.~\cite{Vanasse:2015fph} calculated the triton point charge radius to next-to-next-to leading-order ($\nnlo$) finding good agreement with experiment.  This paper builds upon this work by considering the electric  and magnetic properties of three-nucleon systems in the absence of Coulomb interactions.  In fact the calculation of the general three-nucleon form factor, resulting moments (value at $Q^{2}=0$), and radii for any external current with non-derivative coupling is considered in this work.  This is possible since the form factors for such currents depend on the same integrals but with different constants in front of them. 

In \EFT the charge form factor up to $\nnlo$ can be predicted using four two-body LECs and two three-body LECs encoding interactions between nuclei.  The two-body LECs in this work are fit to the $^{3}S_{1}$ and $^{1}S_{0}$ poles for $N\!N$ scattering and their associated residues, while the three-body LECs are fit to the triton binding energy and the doublet $S$-wave $nd$ scattering length.  In this work Coulomb interactions and isospin breaking from strong interactions are ignored for $\jjvHe$, therefore next-to leading order (NLO) and $\nnlo$ Coulomb and isospin breaking corrections to the three-body force can be ignored~\cite{Vanasse:2014kxa}.  The three-nucleon \EFT magnetic form factor to NLO requires the same LECs as the charge form factor with the exception of the $\nnlo$ energy dependent three-body force.  In addition the NLO magnetic form factor will require an isoscalar and isovector two-body magnetic current.  

The three-nucleon charge form factors are reproduced well using potential model calculations (PMCs)~\cite{Schiavilla:1990zz,Marcucci:1998tb}, whereas the magnetic form factor of $\jjvH$ is reasonably reproduced, but the $\jjvHe$ magnetic form factor poorly describes the first observed diffraction minimum from experiment.  Chiral EFT ($\chi$EFT)~\cite{Piarulli:2012bn} reproduces the three-nucleon charge and magnetic form factors well for $Q\lesssim 3~\mathrm{fm}^{-1}$.  The resulting charge radii, magnetic moments, and magnetic radii from PMCs and $\chi$EFT agree reasonably well with experimental data.\footnote{For a comparison between different methods, including \EFT, for calculating the triton charge radius consult Ref.~\cite{Vanasse:2015fph}.}  \EFT is only valid for momentum transfers of $Q\lesssim 0.7~\mathrm{fm}^{-1}$ and thus cannot directly address the issues observed in PMCs and $\chi$EFT for larger $Q$ values.  However, \EFT can garner insight into the importance of two- and three-body currents.  

As shown in Ref.~\cite{Vanasse:2016umz}, going to the Wigner-SU(4) symmetric limit in which the $N\!N$ scattering lengths and effective ranges for the $^{3}\!S_{1}$ and $^{1}\!S_{0}$ channels are set equal reproduces properties (\emph{e.g.} bound state energy and charge radii) of the three-nucleon systems well within expected errors.  It was also shown that a dual perturbative expansion in \EFT and powers of a Wigner-SU(4) symmetry breaking parameter led to good convergence with experimental data for three-nucleon systems.  Expanding on this, the values of the three-nucleon magnetic moments in the Wigner-SU(4) symmetric limit are calculated in this work.  At LO in this limit the Schmidt-limit~\cite{Schmidt1937} is reproduced in which the magnetic moment of the three nucleon system is given by the magnetic moment of the unpaired nucleon.  It is also demonstrated in the Wigner-SU(4) limit that the expressions for the NLO magnetic moments can be written entirely in terms of LO three-nucleon vertex functions.

This paper is organized as follows.  Section~\ref{sec:twobody} gives the \EFT Lagrangian and all necessary two-body physics, while Sec.~\ref{sec:threebody} reviews relevant properties of the three-body system.  In Sec.~\ref{sec:formfactor} properties of the charge and magnetic form factor in \EFT are derived, and the consequences of Wigner-symmetry on the form factors discussed.  Finally, in Sec.~\ref{sec:results} results are given and conclusions are given in Sec.~\ref{sec:conclusions}.


\section{\label{sec:twobody}Lagrangian and Two-Body System}

\begin{figure}[hbt]
\includegraphics[width=110mm]{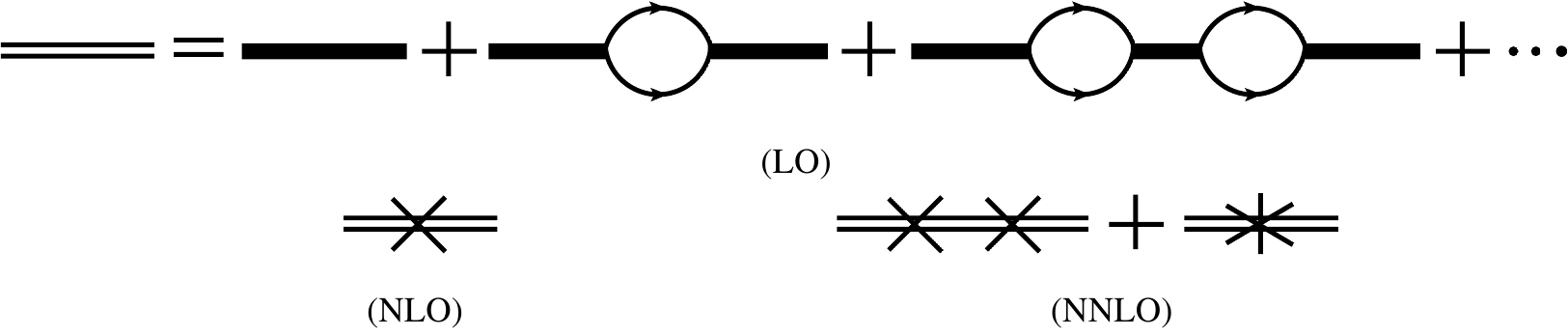}
\caption{The top equation shows the LO dressed spin-triplet dibaryon propagator, which can be solved analytically via a geometric series.  Nucleons are single lines, solid lines are the bare spin-triplet dibaryon propagator $i/\Delta_{t}$, and the double line is the dressed spin-triplet dibaryon.  The cross represents a NLO order effective range insertion from $c_{0t}^{(0)}$ and the star a $\nnlo$ correction from $c_{0t}^{(1)}$.\label{fig:DeutProp}}
\end{figure}

The two-body \EFT Lagrangian is

\begin{align}
\mathcal{L}_{2}=\ &\hat{N}^{\dagger}\left(iD_{0}+\frac{\vect{D}^2}{2M_{N}}\right)\hat{N}+\hat{t}_{i}^{\dagger}\left[\Delta_{t}-c_{0t}\left(iD_{0}+\frac{\vect{D}^{2}}{4M_{N}}+\frac{\gamma_{t}^{2}}{M_{N}}\right)\right]\hat{t}_{i}\\\nonumber
&+\hat{s}_{a}^{\dagger}\left[\Delta_{s}-c_{0s}\left(iD_{0}+\frac{\vect{D}^{2}}{4M_{N}}+\frac{\gamma_{s}^{2}}{M_{N}}\right)\right]\hat{s}_{a}\\\nonumber
&+y_{t}\left[\hat{t}_{i}^{\dagger}\hat{N} ^{T}P_{i}\hat{N} +\mathrm{H.c.}\right]+y_{s}\left[\hat{s}_{a}^{\dagger}\hat{N}^{T}\bar{P}_{a}\hat{N}+\mathrm{H.c.}\right],
\end{align}
where $\hat{t}_{i}$ ($\hat{s}_{a}$) is the spin-triplet (spin-singlet) dibaryon field.  Parameter $y_{t}$ ($y_{s}$) sets the interaction strength between the spin-triplet (spin-singlet) dibaryon and nucleons, while $P_{i}=\frac{1}{\sqrt{8}}\sigma_{2}\sigma_{i}\tau_{2}$ ($\bar{P}_{a}=\frac{1}{\sqrt{8}}\tau_{2}\tau_{a}\sigma_{2}$) projects out the spin-triplet iso-singlet (spin-singlet iso-triplet) combination of nucleons.  The covariant derivative is defined by 
\begin{equation}
D_{\mu}=\partial_{\mu}+i\mathbf{Q}\hat{A}_{\mu},
\end{equation}
where $\hat{A}_{\mu}$ is the photon field, and $\mathbf{Q}$ is the charge operator given by $\mathbf{Q}=(1+\tau_{3})/2$, $\mathbf{Q}=1$, and $\mathbf{Q}=(1+T_{3})$ for the fields $\hat{N}$, $\hat{t}_{i}$, and $\hat{s}_{a}$ respectively.\footnote{$T_{3}$ is the operator for the z-component of isospin.}  $i/\Delta_{t}$ is the bare spin-triplet dibaryon propagator which at LO is dressed by an infinite series of nucleon bubble diagrams as shown in Fig.~\ref{fig:DeutProp}.  This series, a geometric series, yields the LO spin-triplet dibaryon propagator, which receives range corrections from $c_{0t}$ at NLO and $\nnlo$ as shown in Fig.~\ref{fig:DeutProp}.  The resulting parameters of the spin-triplet dibaryon propagator are then fit to give the deuteron pole at LO and its residue at higher orders.  The same procedure can be carried out for the spin-singlet dibaryon propagator with parameters fit to the $^{1}S_{0}$ virtual bound state pole at LO and to its residue at NLO.  This fitting procedure is known as the $Z$-parametrization~\cite{Phillips:1999hh,Griesshammer:2004pe} and has the advantage of giving the correct residue about the poles in the $^{3}S_{1}$ and $^{1}S_{0}$ channels at NLO instead of being approached perturbatively as in the effective range expansion (ERE) parametrization.
Using the $Z$-parametrization gives the coefficients~\cite{Griesshammer:2004pe}
\begin{align}
&y_{t}^{2}=\frac{4\pi}{M_{N}},\quad \Delta_{t}=\gamma_{t}-\mu,\quad c_{0t}^{(n)}=(-1)^{n}(Z_{t}-1)^{n+1}\frac{M_{N}}{2\gamma_{t}},\\\nonumber
&y_{s}^{2}=\frac{4\pi}{M_{N}},\quad \Delta_{s}=\gamma_{s}-\mu,\quad \!c_{0s}^{(n)}=(-1)^{n}(Z_{s}-1)^{n+1}\frac{M_{N}}{2\gamma_{s}},
\end{align}
where $\gamma_{t}=45.7025$~MeV is the deuteron binding momentum, $Z_{t}=1.6908$ is the residue about the deuteron pole, $\gamma_{s}=-7.890$~MeV is the ${}^{1}\!S_{0}$ virtual bound-state momentum, and $Z_{s}=0.9015$ is the residue about the ${}^{1}\!S_{0}$ pole \cite{deSwart:1995ui}.  The scale $\mu$ comes from using dimensional regularization with the power-divergence subtraction scheme~\cite{Kaplan:1998tg,Kaplan:1998we}, and all physical observables do not depend on $\mu$.  Parameter $c_{0t}$ ($c_{0s}$) is split up into contributions $c_{0t}^{(n)}$ ($c_{0s}^{(n)}$) at each order to ensure the pole position is fixed and has the correct residue.  The resulting spin-triplet (spin-singlet) dibaryon in the $Z$-parametrization up to $\nnlo$ is given by
\begin{align}
&iD_{\{t,s\}}^{\mathrm{NNLO}}(p_{0},\vect{p})=\frac{i}{\gamma_{\{t,s\}}-\sqrt{\frac{\vect{p}^{2}}{4}-M_{N}p_{0}-i\epsilon}}\\\nonumber
&\times\left[\underbrace{\srutTypeOne{.5cm} 1}_{\mathrm{LO}}+\underbrace{\frac{Z_{\{t,s\}}-1}{2\gamma_{\{t,s\}}}\left(\gamma_{\{t,s\}}+\sqrt{\frac{\vect{p}^{2}}{4}-M_{N}p_{0}-i\epsilon}\right)}_{\mathrm{NLO}}\right.\\\nonumber
&\left.\hspace{.5cm}+\underbrace{\left(\frac{Z_{\{t,s\}}-1}{2\gamma_{\{t,s\}}}\right)^{2}\left(\frac{\vect{p}^{2}}{4}-M_{N}p_{0}-\gamma_{\{t,s\}}^{2}\right)}_{\mathrm{NNLO}}+\cdots\right].\\\nonumber
\end{align}
%

LO interactions between nucleons and the magnetic field at the one-body level are given by the Lagrangian
\begin{equation}
\label{eq:1BLagMag}
\mathcal{L}_{1,0}^{mag}=\frac{e}{2M_{N}}\hat{N}^{\dagger}(\kappa_{0}+\kappa_{1}\tau_{3})\vect{\sigma}\cdot\mathbf{B}\hat{N},
\end{equation}
where $\kappa_{0}=0.4399$ is the isoscalar magnetic moment of the nucleon and $\kappa_{1}=2.3529$ is the isovector magnetic moment of the nucleon in nuclear magnetons.  At NLO there are two two-body magnetic currents, $L_{1}$~\cite{Chen:1999tn,Beane:2000fi} and $L_{2}$~\cite{Kaplan:1998sz,Chen:1999tn} given by the Lagrangian
\begin{equation}
\label{eq:2BLagMag}
\mathcal{L}_{2}^{mag}=\left(e\frac{L_{1}}{2}\hat{t}^{j\dagger}\hat{s}_{3}\mathbf{B}_{j}+\mathrm{H.c.}\right)-e\frac{L_{2}}{2}i\epsilon^{ijk}\hat{t}_{i}^{\dagger}\hat{t}_{j}\mathbf{B}_{k}.
\end{equation}

In the three-body system there will be a LO three-body force~\cite{Bedaque:1999ve} with non-derivative coupling, which receives corrections at higher orders to avoid refitting.  At $\nnlo$ a new energy dependent three-body force is required in \EFT~\cite{Bedaque:2002yg}.  These three-body forces are easily represented by the introduction of an interaction between $\hat{\psi}$~\cite{Bedaque:2002yg,Vanasse:2015fph}, dibaryons, and nucleons via the Lagrangian
\begin{align}
{\mathcal{L}}_{3}=&\hat{\psi}^{\dagger}\left[\Omega-h_{2}(\Lambda)\left(iD_{0}+\frac{\vect{D}^{2}}{6M_{N}}+\frac{\gamma_{t}^{2}}{M_{N}}\right)\right]\hat{\psi}+\sum_{n=0}^{\infty}\left[\omega^{(n)}_{t0}\hat{\psi}^{\dagger}\sigma_{i}\hat{N}\hat{t}_{i}-\omega^{(n)}_{s0}\hat{\psi}^{\dagger}\tau_{a}\hat{N}\hat{s}_{a}\right]\\\nonumber
&+\mathrm{H.c.}, 
\end{align}
where $\hat{\psi}$ is a three-nucleon iso-doublet field containing $\jjvH$ and $\jjvHe$.  The $\nnlo$ energy dependent three-body force term is given by
\begin{equation}
\label{eq:H2}
\widehat{H}_{2}=-\frac{3(\omega^{(0)}_{t0})^{2}}{\pi\Omega^{2}M_{N}}h_{2}(\Lambda)=-\frac{3(\omega^{(0)}_{s0})^{2}}{\pi\Omega^{2}M_{N}}h_{2}(\Lambda)=-\frac{3\omega^{(0)}_{t0}\omega^{(0)}_{s0}}{\pi\Omega^{2}M_{N}}h_{2}(\Lambda).
\end{equation}
For further details of three-body forces and how they are fit consult Ref.~\cite{Vanasse:2015fph}.


\section{\label{sec:threebody}Three-Body System}

\begin{figure}[hbt]
\includegraphics[width=100mm]{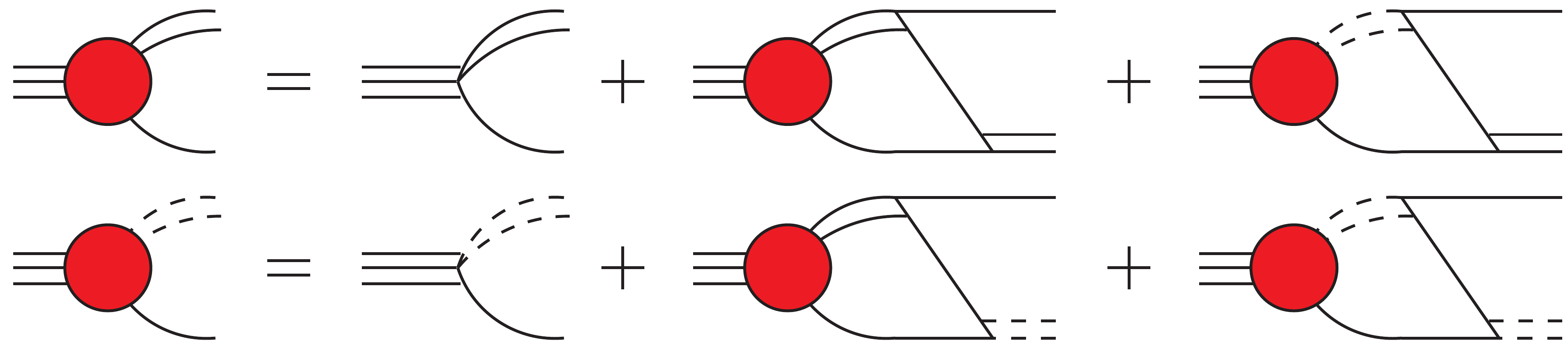}
\caption{The coupled-channel integral equations for the LO three-nucleon vertex function, where the triple line is the three-nucleon system, and the filled circle is the LO three-nucleon vertex function.\label{fig:GirrLO}}
\end{figure}

Detailed methods for calculating the three-nucleon vertex function can be found in Ref.~\cite{Vanasse:2015fph} and a brief review of them, in order that this work is relatively self contained, is given below.  The LO three-nucleon vertex function is  the solution of an integral equation represented by the diagrams of Fig.~\ref{fig:GirrLO}.
Double dashed lines are spin-singlet dibaryons and the triple lines three-nucleon fields.  In cluster-configuration (c.c.)~space~\cite{Griesshammer:2004pe} the LO three-nucleon vertex function is given by the integral equation
\begin{equation}
\label{eq:LOGirr}
\Gb_{0}(E,p)=\mathbf{\widetilde{B}}_{0}+\mathbf{K}_{0}(q,p,E)\otimes\Gb_{0}(E,q),
\end{equation}
where $\Gb_{0}(E,p)$ is a c.c.~space vector given by

\begin{equation}
\label{eq:GDef}
\boldsymbol{\mathcal{G}}_{0}(E,p)=\left(
\begin{array}{c}
\G_{0,\psi\to Nt}(E,p)\\
\G_{0,\psi\to Ns}(E,p)
\end{array}\right),
\end{equation}
and the inhomogeneous term $\mathbf{\widetilde{B}}_{0}$ is a c.c.~space vector given by
\begin{equation}
\label{eq:BtildeDef}
\mathbf{\widetilde{B}}_{0}=
\left(\!\!\begin{array}{r}
1\\[-.2cm]
-1
\end{array}\right).
\end{equation}
$\G_{0,\psi\to Nt}(E,p)$ ($\G_{0,\psi\to Ns}(E,p)$) is the three-nucleon vertex function for a three-nucleon system going to a nucleon and deuteron (nucleon and spin-singlet dibaryon).  The kernel of Eq.~(\ref{eq:LOGirr}) is a c.c.~space matrix given by
\begin{equation}
\hspace{-.5cm}\mathbf{K}_{0}(q,p,E)=\mathbf{R}_{0}(q,p,E)\,\mathbf{D}^{(0)}\!\!\left(E-\frac{q^{2}}{2M_{N}},\vect{q}\right),
\end{equation}
where
\begin{align}
\mathbf{R}_{0}(q,p,E)=-\frac{2\pi}{qp}Q_{0}\left(\frac{q^{2}+p^{2}-M_{N}E-i\epsilon}{qp}\right)\left(\!\!\!
\begin{array}{rr}
1 & -3 \\
-3 & 1
\end{array}\!\right),
\end{align}
matrix multiplies
\begin{equation}
\label{eq:DibMatrix}
\mathbf{D}^{(0)}(E,\vect{q})=
\left(
\begin{array}{cc}
D_{t}^{(0)}(E,\vect{q}) & 0 \\
0&D_{s}^{(0)}(E,\vect{q})  
\end{array}\right),
\end{equation}
which is a matrix of LO dibaryon propagators.  $Q_{0}(a)$ is a Legendre function of the second kind defined as
\begin{equation}
Q_{0}(a)=\frac{1}{2}\ln\left(\frac{1+a}{1-a}\right),
\end{equation}
and the ``$\otimes$" notation is defined by
\begin{equation*}
\label{eq:otimes}
A(q)\!\otimes\! B(q)=\frac{1}{2\pi^{2}}\int_{0}^{\Lambda}dqq^{2}A(q)B(q).
\end{equation*}

\begin{figure}[hbt]
\includegraphics[width=100mm]{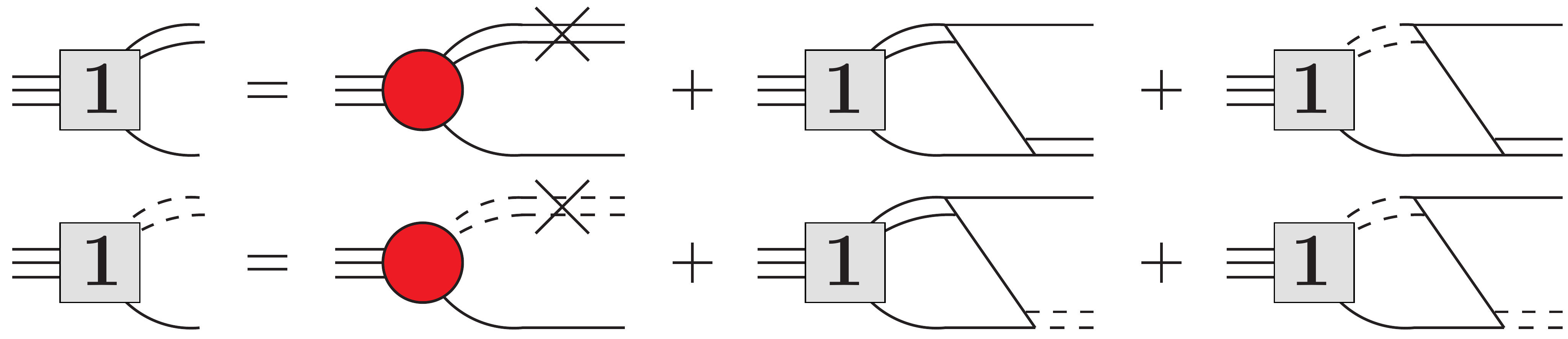}
\caption{The coupled-channel integral equations for the NLO correction to the three-nucleon vertex function.\label{fig:GirrNLO}}
\end{figure}

\begin{figure}[hbt]
\includegraphics[width=120mm]{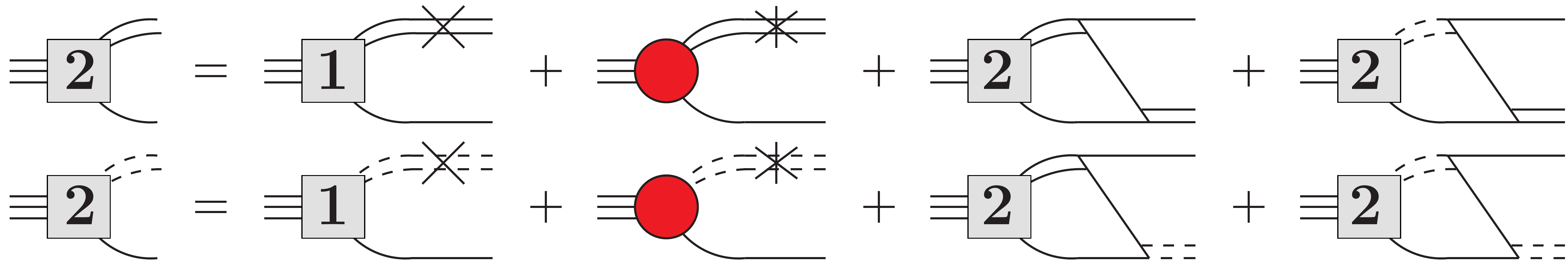}
\caption{The coupled-channel integral equations for the $\nnlo$ correction to the three-nucleon vertex function.\label{fig:GirrNNLO}}
\end{figure}

The NLO and $\nnlo$ three-nucleon vertex functions are given by integral equations represented in Figs.~\ref{fig:GirrNLO} and \ref{fig:GirrNNLO} respectively.
In c.c.~space the NLO three-nucleon vertex function is
\begin{equation}
\Gb_{1}(E,p)=\mathbf{R}_{1}\!\!\left(E-\frac{\vect{p}^{2}}{2M_{N}},\vect{p}\right)\Gb_{0}(E,p)+\mathbf{K}_{0}(q,p,E)\otimes\Gb_{1}(E,q),
\end{equation}
where $\mathbf{R}_{1}(p_{0},\vect{p})$ is a c.c.~space matrix defined by
\begin{equation}
\mathbf{R}_{1}(p_{0},\vect{p})=
\left(\begin{array}{cc}
\frac{Z_{t}-1}{2\gamma_{t}}\left(\gamma_{t}+\sqrt{\frac{1}{4}\vect{p}^{2}-M_{N}p_{0}-i\epsilon}\,\right) & 0\\
0 &\frac{Z_{s}-1}{2\gamma_{s}}\left(\gamma_{s}+\sqrt{\frac{1}{4}\vect{p}^{2}-M_{N}p_{0}-i\epsilon}\,\right)
\end{array}\right).
\end{equation}
In c.c.~space the $\nnlo$ three-nucleon vertex function is given by
\begin{equation}
\Gb_{2}(E,p)=\mathbf{R}_{1}\left(E-\frac{\vect{p}^{2}}{2M_{N}},\vect{p}\right)\Big[\Gb_{1}(E,p)-\mathbf{c}_{1}\Gb_{0}(E,p)\Big]+\mathbf{K}_{0}(q,p,E)\otimes\Gb_{2}(E,q),
\end{equation}
where 
\begin{equation}
\mathbf{c}_{1}=\left(\begin{array}{cc}
Z_{t}-1 & 0\\
0 & Z_{s}-1
\end{array}\right),
\end{equation}
is a c.c.~space matrix.

To properly normalize the three-nucleon vertex function the three-nucleon wavefunction renormalization is needed, which is obtained by calculating the residue about the three-nucleon propagator pole.  This pole is fixed to the triton binding energy $B=E_{\jjvH}$, $E_{\jjvH}=-8.48$~MeV~\cite{Wapstra:1985zz}, by appropriate tuning of three-body forces.  Further details of how this is done can be seen in Ref.~\cite{Vanasse:2015fph}.  The resulting three-nucleon wavefunction renormalization up to and including $\nnlo$ is given by
\begin{align}
&Z_{\psi}=\frac{\pi}{\Sigma_{0}'(B)}\left[\underbrace{\vphantom{\frac{\Sigma_{1}'(B)}{\Sigma_{0}'(B)}}1}_{\mathrm{LO}}-\underbrace{\frac{\Sigma_{1}'(B)}{\Sigma_{0}'(B)}}_{\mathrm{NLO}}\right.\\\nonumber
&\hspace{2cm}\left.-\underbrace{\left\{\frac{\Sigma_{2}'(B)}{\Sigma_{0}'(B)}-\left(\frac{\Sigma_{1}'(B)}{\Sigma_{0}'(B)}\right)^{2}+\frac{4}{3}M_{N}\widehat{H}_{2}\Sigma_{0}(B)\left(\frac{\Sigma_{0}(B)}{\Sigma_{0}'(B)}-B-\frac{\gamma_{t}^{2}}{M_{N}}\right)\!\!\right\}}_{\nnlo}+\cdots\right],
\end{align}
where the $\Sigma_{n}(E)$ functions are defined by
\begin{equation}
\Sigma_{n}(E)=-\pi\mathrm{Tr}\left[\mathbf{D}^{(0)}\!\!\left(E-\frac{q^{2}}{2M_{N}},q\right)\otimes\Gb_{n}(E,q)\right],
\end{equation}
and $\hat{H}_{2}$ is the energy dependent $\nnlo$ three-body force~\cite{Vanasse:2015fph,Bedaque:2002yg} from Eq.~(\ref{eq:H2}).  Taking the square root of $Z_{\psi}$ and expanding, the properly renormalized LO three-nucleon vertex function is given by
\begin{equation}
\boldsymbol{\Gamma}_{0}(p)=\sqrt{Z_{\psi}^{\mathrm{LO}}}\Gb_{0}(B,p),
\end{equation}
the properly renormalized NLO correction to the three-nucleon vertex function by
\begin{equation}
\boldsymbol{\Gamma}_{1}(p)=\sqrt{Z^{\mathrm{LO}}_{\psi}}\left[\Gb_{1}(B,p)-\frac{1}{2}\frac{\Sigma_{1}'(B)}{\Sigma_{0}'(B)}\Gb_{0}(B,p)\right],
\end{equation}
and the properly renormalized $\nnlo$ correction to the three-nucleon vertex function by
\begin{align}
\boldsymbol{\Gamma}_{2}(p)&=\sqrt{Z^{\mathrm{LO}}_{\psi}}\left[\Gb_{2}(B,p)-\frac{1}{2}\frac{\Sigma_{1}'(B)}{\Sigma_{0}'(B)}\Gb_{1}(B,p)\right.\\\nonumber
&\hspace{2cm}\left.+\left\{\frac{3}{8}\left(\frac{\Sigma_{1}'(B)}{\Sigma_{0}'(B)}\right)^{2}-\frac{1}{2}\frac{\Sigma_{2}'(B)}{\Sigma_{0}'(B)}-\frac{2}{3}M_{N}\widehat{H}_{2}\frac{\Sigma_{0}^{2}(B)}{\Sigma_{0}'(B)}\right\}\Gb_{0}(B,p)\right],
\end{align}
where
\begin{equation}
Z^{\mathrm{LO}}_{\psi}=\frac{\pi}{\Sigma_{0}'(B)}.
\end{equation}


\section{\label{sec:formfactor}Charge and Magnetic Form Factors}

\subsection{Charge and Magnetic Moments}

\begin{figure}[hbt]
\includegraphics[width=100mm]{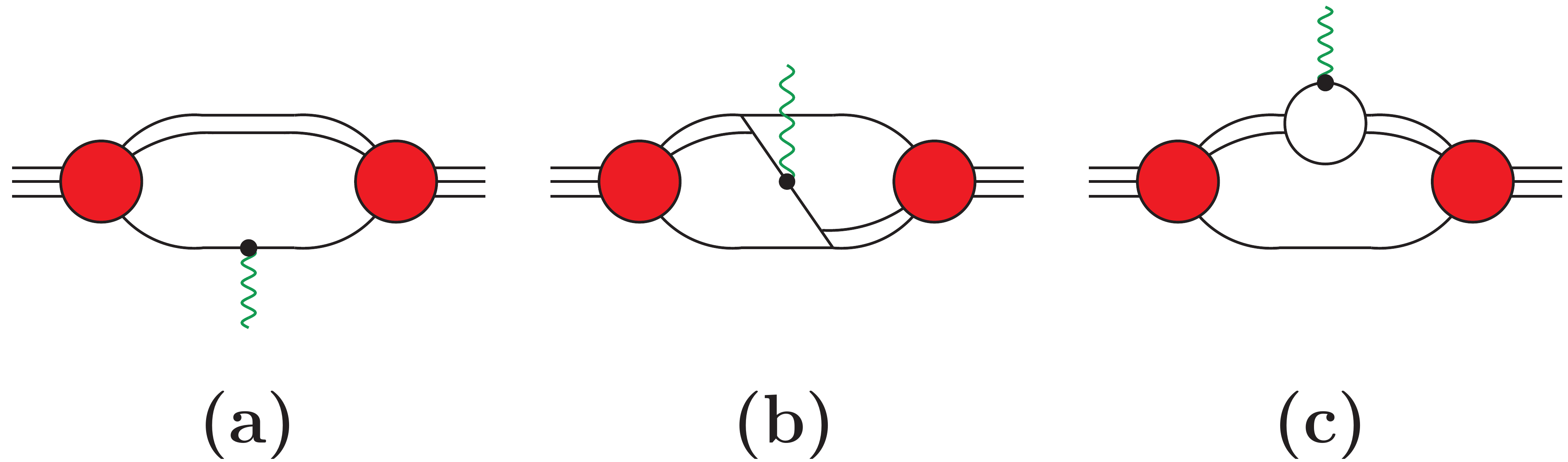}
\caption{Diagrams for the LO three-nucleon charge and magnetic form factor.  All wavy green lines represent either a magnetic or $\hat{A}_{0}$ photon and the black dot their respective coupling to the nucleons.\label{fig:FormFactorLO}}
\end{figure}

In Ref.~\cite{Vanasse:2015fph} the charge form factor of the triton was calculated to $\nnlo$ in \EFT.  Calculating the $\jjvHe$ charge form factor, $\jjvH$ magnetic form factor, and the $\jjvHe$ magnetic form factor in the absence of Coulomb interactions is essentially the same calculation as the $\jjvH$ charge form factor.  The only difference between these calculations are the coefficients that appear in front of the same integrals.  Both charge and magnetic form factors at LO are given by the sum of diagrams in Fig.~\ref{fig:FormFactorLO}, where all photons are either minimally coupled $\hat{A}_{0}$ photons or magnetically coupled from Eq.~(\ref{eq:1BLagMag}).
Form factors are calculated in the Breit frame in which the photon only imparts momentum $\vect{Q}$ but no energy on the three-nucleon system, and all form factors are only functions of $Q^{2}$.  Using the work of Ref.~\cite{Vanasse:2015fph} the LO ``generic" form factor in the limit $Q^{2}=0$ is given by
\begin{align}
\label{eq:LOGenForm}
&F_{0}(0)=2\pi M_{N}\left(\widetilde{\boldsymbol{\Gamma}}_{0}(q)\right)^{T}\otimes\left\{\frac{\pi}{2}\frac{\delta(q-\ell)}{q^{2}\sqrt{\frac{3}{4}q^{2}-M_{N}B}}
\left(\begin{array}{cc}
c_{11}+a_{11} & c_{12} \\
c_{21} & c_{22}+a_{22}
\end{array}\right)\right.\\\nonumber
&\hspace{1cm}\left.+\frac{1}{q^{2}\ell^{2}-(q^{2}+\ell^{2}-M_{N}B)^{2}}
\left(\!\!\begin{array}{rr}
b_{11}-2a_{11} & b_{12}+3(a_{11}+a_{22}) \\
b_{21}+3(a_{11}+a_{22}) & b_{22}-2a_{22}
\end{array}\!\right)\right\}\otimes\widetilde{\boldsymbol{\Gamma}}_{0}(\ell),
\end{align}
where the c.c.~space vector function $\widetilde{\boldsymbol{\Gamma}}_{n}(q)$ is
\begin{equation}
\widetilde{\boldsymbol{\Gamma}}_{n}(q)=\mathbf{D}^{(0)}\left(B-\frac{q^{2}}{2M_{N}},\vect{q}\right)\boldsymbol{\Gamma}_{n}(q),
\end{equation}
and $n=0,1,2,\cdots$.  The coefficients $a_{11}$ and $a_{22}$ come from the c.c.~space matrix of diagram Fig.~\ref{fig:FormFactorLO}(a), the coefficients $b_{11}$, $b_{12}$, $b_{21}$, and $b_{22}$ from the c.c.~space matrix of diagram Fig.~\ref{fig:FormFactorLO}(b), and the coefficients $c_{11}$, $c_{12}$, $c_{21}$, and $c_{22}$ from the c.c.~space matrix of diagram Fig.~\ref{fig:FormFactorLO}(c).  The only difference between the LO magnetic and charge form factors for $\jjvH$ and $\jjvHe$ are the values of these coefficients shown in Table~\ref{tab:LOvalues} for each.
\begin{table}
\begin{tabular}{|c|c|c|c|c|c|c|c|c|c|c|}
\hline
Form factor & $a_{11}$ & $a_{22}$ & $b_{11}$ & $b_{12}$ & $b_{21}$ & $b_{22}$ & $c_{11}$ & $c_{12}$ & $c_{21}$ & $c_{22}$ \\
\hline
$F_{C}^{\jjvH}(Q^{2})$ & 0 & $\frac{2}{3}$ & -1 & 1 & 1 & $\frac{1}{3}$ & 1 & 0 & 0 & $\frac{1}{3}$  \\\hline
$F_{C}^{\jjvHe}(Q^{2})$ & 1 & $\frac{1}{3}$ & 0 & 2 & 2 & -$\frac{4}{3}$ & 1 & 0 & 0 & $\frac{5}{3}$ \\\hline
$F_{M}^{\jjvH}(Q^{2})$ & $\frac{\kappa_{1}\!-\!\kappa_{0}}{3}$ & \small$\kappa_{0}\!+\!\frac{1}{3}\kappa_{1}$ & $-\frac{5(\kappa_{0}+\kappa_{1})}{3}$ & $\kappa_{0}-\frac{1}{3}\kappa_{1}$ & $\kappa_{0}-\frac{1}{3}\kappa_{1}$ & $\kappa_{0}-\frac{5}{3}\kappa_{1}$ & $\frac{4}{3}\kappa_{0}$ & $-\frac{2}{3}\kappa_{1}$ & $-\frac{2}{3}\kappa_{1}$ & 0 \\\hline
$F_{M}^{\jjvHe}(Q^{2})$ & $-\frac{\kappa_{0}\!+\!\kappa_{1}}{3}$ & $\kappa_{0}\!-\!\frac{1}{3}\kappa_{1}$ & $-\frac{5(\kappa_{0}-\kappa_{1})}{3}$ & $\kappa_{0}+\frac{1}{3}\kappa_{1}$ & $\kappa_{0}+\frac{1}{3}\kappa_{1}$ & $\kappa_{0}+\frac{5}{3}\kappa_{1}$ & $\frac{4}{3}\kappa_{0}$ & $\frac{2}{3}\kappa_{1}$ & $\frac{2}{3}\kappa_{1}$ & 0 \\\hline
\end{tabular}
\caption{\label{tab:LOvalues}Values of coefficients for the LO $\jjvH$ and $\jjvHe$ magnetic and charge form factors.  Note factors of $\frac{e}{2M_{N}}$ have been removed from the magnetic coefficients since the magnetic moments are given in units of nuclear magnetons.}
\end{table}
Further details of how these coefficients are obtained are given in Appendix~\ref{app:chi}. 

Choosing the coefficients for the triton charge form factor gives
\begin{align}
\label{eq:F0triton}
&F_{0}(0)=2\pi M_{N}\left(\widetilde{\boldsymbol{\Gamma}}_{0}(q)\right)^{T}\otimes\left\{\frac{\pi}{2}\frac{\delta(q-\ell)}{q^{2}\sqrt{\frac{3}{4}q^{2}-M_{N}B}}
\left(\begin{array}{cc}
1 & 0 \\
0 & 1
\end{array}\right)\right.\\\nonumber
&\hspace{1cm}\left.-\frac{1}{q^{2}\ell^{2}-(q^{2}+\ell^{2}-M_{N}B)^{2}}
\left(\!\!\begin{array}{rr}
1 & -3 \\
-3 & 1
\end{array}\!\right)\right\}\otimes\widetilde{\boldsymbol{\Gamma}}_{0}(\ell).
\end{align}
This expression is the same as the normalization condition in Ref.~\cite{Konig:2011yq}, and therefore it follows automatically that $F_{0}(0)=1$ for the triton charge form factor.  Plugging in the $\jjvHe$ charge form factor coefficients gives two times Eq.~(\ref{eq:F0triton}), and hence $F_{0}(0)=2$ for the $\jjvHe$ charge form factor.\footnote{Conventionally charge form factors are defined such that $F(Q^{2}=0)=1$.}

\begin{figure}[hbt]
\includegraphics[width=100mm]{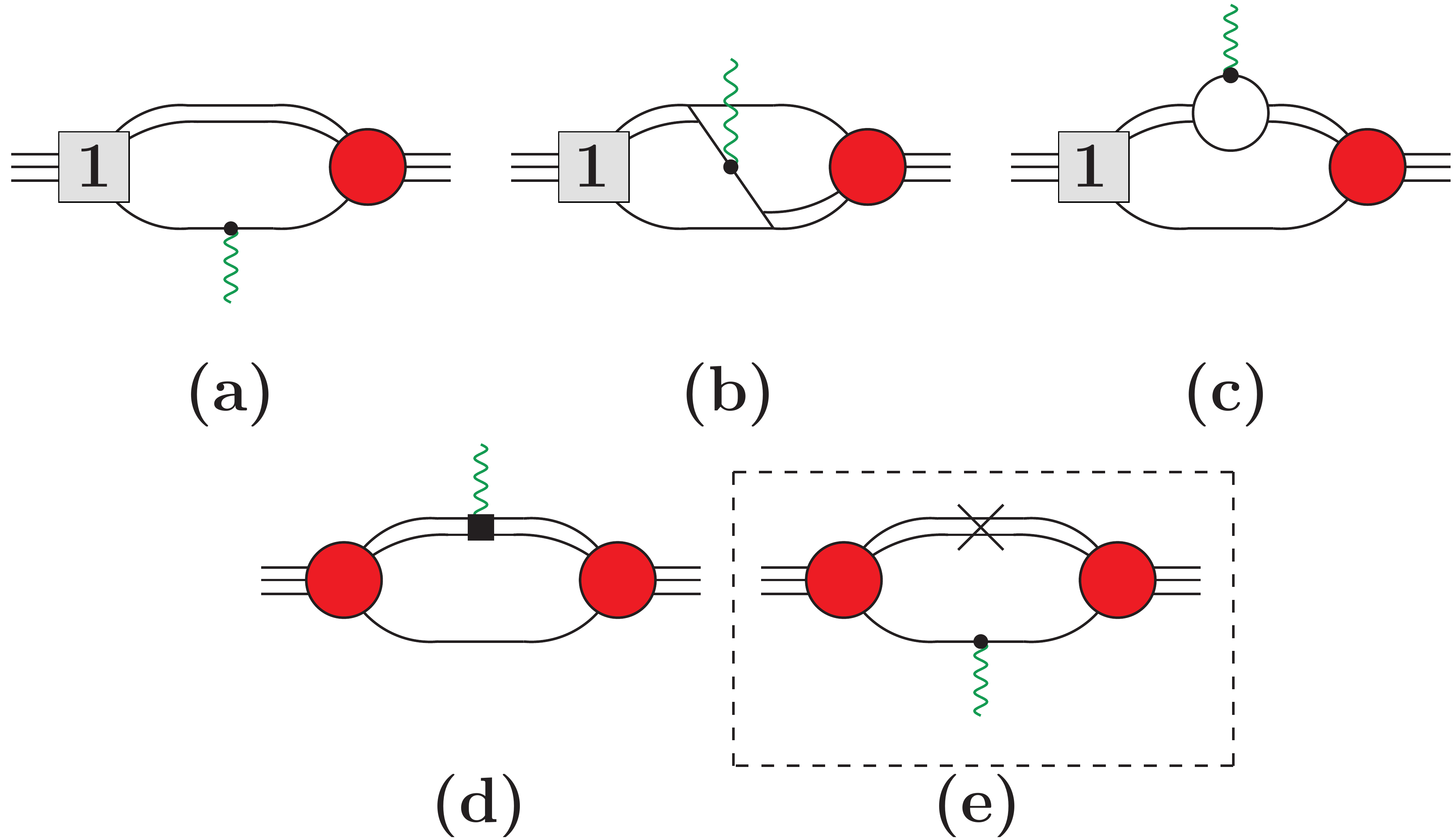}
\caption{Diagrams for the NLO correction to the three-nucleon charge and magnetic form factors.  The boxed diagram is subtracted to avoid double counting.  For the charge form factor diagram (d) comes from gauging the dibaryon kinetic term and for the magnetic form factor from the $L_{1}$ and $L_{2}$ contact terms given in Eq.~(\ref{eq:2BLagMag}).  Diagrams related by time reversal symmetry are not shown.\label{fig:FormFactorNLO}}
\end{figure}

The NLO correction to the charge and magnetic form factors is given by the diagrams in Fig.~\ref{fig:FormFactorNLO}.  
Diagram-(d) for charge form factors comes from gauging the dibaryon kinetic term and for the magnetic form factor comes from the $L_{1}$ and $L_{2}$ term of Eq.~(\ref{eq:2BLagMag}).  Not shown in Fig.~\ref{fig:FormFactorNLO} are diagrams related by time reversal symmetry.  Diagram-(e) in the dashed box is subtracted from the other diagrams to avoid double counting from diagram-(a) and its time reversed version.  The NLO correction to the ``generic" form factor in the limit $Q^{2}=0$ is
\begin{align}
\label{eq:NLOform}
&F_{1}(0)=2\pi M_{N}\left(\widetilde{\boldsymbol{\Gamma}}_{1}(q)\right)^{T}\otimes\left\{\frac{\pi}{2}\frac{\delta(q-\ell)}{q^{2}\sqrt{\frac{3}{4}q^{2}-M_{N}B_{0}}}
\left(\begin{array}{cc}
c_{11}+a_{11} & c_{12} \\
c_{21} & c_{22}+a_{22}
\end{array}\right)\right.\\\nonumber
&\hspace{1cm}\left.+\frac{1}{q^{2}\ell^{2}-(q^{2}+\ell^{2}-M_{N}B_{0})^{2}}
\left(\!\!\begin{array}{rr}
b_{11}-2a_{11} & b_{12}+3(a_{11}+a_{22}) \\
b_{21}+3(a_{11}+a_{22}) & b_{22}-2a_{22}
\end{array}\!\right)\right\}\otimes\widetilde{\boldsymbol{\Gamma}}_{0}(\ell)\\\nonumber
&+2\pi M_{N}\left(\widetilde{\boldsymbol{\Gamma}}_{0}(q)\right)^{T}\otimes\left\{\frac{\pi}{2}\frac{\delta(q-\ell)}{q^{2}\sqrt{\frac{3}{4}q^{2}-M_{N}B_{0}}}
\left(\begin{array}{cc}
c_{11}+a_{11} & c_{12} \\
c_{21} & c_{22}+a_{22}
\end{array}\right)\right.\\\nonumber
&\hspace{1cm}\left.+\frac{1}{q^{2}\ell^{2}-(q^{2}+\ell^{2}-M_{N}B_{0})^{2}}
\left(\!\!\begin{array}{rr}
b_{11}-2a_{11} & b_{12}+3(a_{11}+a_{22}) \\
b_{21}+3(a_{11}+a_{22}) & b_{22}-2a_{22}
\end{array}\!\right)\right\}\otimes\widetilde{\boldsymbol{\Gamma}}_{1}(\ell)\\\nonumber
&-4\pi M_{N}\left(\widetilde{\boldsymbol{\Gamma}}_{0}(q)\right)^{T}\otimes\left\{\frac{\pi}{2}\frac{\delta(q-\ell)}{q^{2}}
\left(\begin{array}{cc}
\frac{c_{0t}^{(0)}}{M_{N}}a_{11}+d_{11} & d_{12} \\
d_{21} & \frac{c_{0s}^{(0)}}{M_{N}}a_{22}+d_{22}
\end{array}\right)\right\}\otimes\widetilde{\boldsymbol{\Gamma}}_{0}(\ell),
\end{align}
where the coefficients $d_{11}$,$d_{12}$,$d_{21}$, and $d_{22}$ are from the NLO c.c.~space matrix for diagram Fig.~\ref{fig:FormFactorNLO}(d) and are shown in Table~\ref{tab:dvalues}.  Again the derivation of these coefficients is given in Appendix~\ref{app:chi}.  For $F_{1}(0)$ the first two terms simply come from replacing $\widetilde{\boldsymbol{\Gamma}}_{0}(q)$ by $\widetilde{\boldsymbol{\Gamma}}_{1}(q)$ in Eq.~(\ref{eq:LOGenForm}).  The last term of $F_{1}(0)$ has NLO corrections from diagrams (a),(d), and (e) of Fig.~\ref{fig:FormFactorNLO}.  For the three-nucleon charge form factor $F_{1}(0)=0$ as a consequence of gauge symmetry.
\begin{table}
\begin{tabular}{|c|c|c|c|c|}
\hline
Form factor & $d_{11}$ & $d_{12}$ & $d_{21}$ & $d_{22}$  \\\hline
$F_{C}^{\jjvH}(Q^{2})$ & $c_{0t}^{(0)}/M_{N}$ & 0 & 0 & $\frac{1}{3} c_{0s}^{(0)}/M_{N}$  \\\hline
$F_{C}^{\jjvHe}(Q^{2})$ & $c_{0t}^{(0)}/M_{N}$ & 0 & 0 & $\frac{5}{3} c_{0s}^{(0)}/M_{N}$  \\\hline
$F_{M}^{\jjvH}(Q^{2})$ &$-\frac{2}{3}L_{2}$ & $\frac{1}{3}L_{1}$ & $\frac{1}{3}L_{1}$ & 0 \\\hline
$F_{M}^{\jjvHe}(Q^{2})$ &$-\frac{2}{3}L_{2}$ & $-\frac{1}{3}L_{1}$ & $-\frac{1}{3}L_{1}$ & 0 \\\hline
\end{tabular}
\caption{Values of coefficients for the NLO corrections to (d)-type diagrams for the $\jjvH$ and $\jjvHe$ magnetic and charge form factors.  Note factors of $\frac{e}{2M_{N}}$ have been removed from the magnetic coefficients since the magnetic moments are given in units of nuclear magnetons.\label{tab:dvalues}}
\end{table}

\subsection{Charge and Magnetic Radius}

In general the form factor can be expanded in powers of $Q^{2}$ yielding
\begin{equation}
\label{eq:generic}
F_{X}^{{}^{A}Z}(Q^{2})=f^{^{A}Z}_{X}\left(1-\frac{1}{6}\left<\delta r_{X}^{2}\right>^{{}^{A}Z}Q^{2}+\cdots\right),
\end{equation}
where $X=C$ ($X=M$) for the charge (magnetic) form factor, and ${}^{A}Z=\jjvH$ or $\jjvHe$.  $f^{^{A}Z}_{C}$ ($f^{^{A}Z}_{M}$) is the charge (magnetic moment) of the three-nucleon system, and $\left<\delta r_{C}^{2}\right>^{{}^{A}Z}$ ($\left<\delta r_{M}^{2}\right>^{{}^{A}Z}$) is the point charge (magnetic) radius of the three-nucleon system.  Higher order terms in $Q^{2}$ are not considered in this work, because for values of $Q^{2}$ for which \EFT is valid form factors are dominated by the constant and $Q^{2}$ pieces.  Methods for calculating the form factor with all powers of $Q^{2}$ can be seen in Refs.~\cite{Vanasse:2015fph,Hagen:2013xga}. 

The coefficient of the $Q^{2}$ contribution to the ``generic" form factor to any order up to $\nnlo$ from type (a) diagrams is given by
\begin{align}
\label{eq:FLOa}
&\frac{1}{2}\frac{\partial^{2}}{\partial Q^{2}}F_{n}^{(a)}(Q^{2})\Big{|}_{Q^{2}=0}=Z_{\psi}^{\mathrm{LO}}\sum_{i,j=0}^{i+j\leq n}\left\{\widetilde{\Gb}_{i}^{T}(p)\otimes \boldsymbol{\mathcal{A}}_{n-i-j}(p,k)\otimes\widetilde{\Gb}_{j}(k)\right.\\\nonumber
&\hspace{8cm}\left.+2\widetilde{\Gb}_{i}^{T}(p)\otimes \boldsymbol{\mathcal{A}}_{n-i}(p)\delta_{j0}+\mathcal{A}_{n}\delta_{i0}\delta_{j0}\right\},
\end{align}
where the subscripts denote the order of the term in \EFT.  $\boldsymbol{\mathcal{A}}_{n}(p,k)$ is a c.c.~space matrix, $\boldsymbol{\mathcal{A}}_{n}(p)$ is a c.c.~space vector, and $\mathcal{A}_{n}$ is a c.c.~space scalar.  The detailed form of these functions is given in Appendix~\ref{app:QExpansion} and they all depend on the coefficients $a_{11}$ and $a_{22}$.  Note that the NLO diagram-(e) of Fig.~\ref{fig:FormFactorNLO} is absorbed into the NLO expression for diagram-(a)~\cite{Vanasse:2015fph}.  The c.c.~space vector $\widetilde{\Gb}_{n}(p)$ is defined by
\begin{equation}  
\widetilde{\Gb}_{n}(p)=\mathbf{D}^{(0)}\left(B-\frac{p^{2}}{2M_{N}},\vect{p}\right)\Gb_{n}(B,	p).
\end{equation}
Type-(b) diagrams to any order up to $\nnlo$ give a $Q^{2}$ contribution of
\begin{align}
\label{eq:FLOb}
&\frac{1}{2}\frac{\partial^{2}}{\partial Q^{2}}F_{n}^{(b)}(Q^{2})\Big{|}_{Q^{2}=0}=Z_{\psi}^{\mathrm{LO}}\sum_{i=0}^{n}\widetilde{\Gb}_{i}^{T}(p)\otimes \boldsymbol{\mathcal{B}}_{0}(p,k)\otimes\widetilde{\Gb}_{n-i}(k),
\end{align}
where $\boldsymbol{\mathcal{B}}_{0}(p,k)$ is a c.c.~space matrix given in Appendix~\ref{app:QExpansion}.  Functions $\boldsymbol{\mathcal{B}}_{n}(p,k)$ for $n\geq 1$ do not exist. The $Q^{2}$ contribution from type-(c) diagrams to any order up to $\nnlo$ gives
\begin{align}
\label{eq:FLOc}
&\frac{1}{2}\frac{\partial^{2}}{\partial Q^{2}}F_{n}^{(c)}(Q^{2})\Big{|}_{Q^{2}=0}=Z_{\psi}^{\mathrm{LO}}\sum_{i,j=0}^{i+j\leq n}\left\{\widetilde{\Gb}_{i}^{T}(p)\otimes \boldsymbol{\mathcal{C}}_{n-i-j}(p,k)\otimes\widetilde{\Gb}_{j}(k)+\boldsymbol{\mathcal{C}}_{n-i}(k)\otimes\widetilde{\Gb}_{i}(k)\delta_{j0}\right\},
\end{align}
where $\boldsymbol{\mathcal{C}}_{n}(p,k)$ is c.c.~space matrix and $\boldsymbol{\mathcal{C}}_{n}(k)$ is a c.c.~space vector both given in Appendix~\ref{app:QExpansion}.  Finally, the $Q^{2}$ contribution from type-(d) diagrams to any order up to $\nnlo$ gives	
\begin{align}
&\frac{1}{2}\frac{\partial^{2}}{\partial Q^{2}}F_{n}^{(d)}(Q^{2})\Big{|}_{Q^{2}=0}=Z_{\psi}^{\mathrm{LO}}\sum_{i,j=0}^{i+j\leq n-1}\left\{\widetilde{\Gb}_{i}^{T}(p)\otimes \boldsymbol{\mathfrak{D}}_{n-i-j}(p,k)\otimes\widetilde{\Gb}_{j}(k)+\boldsymbol{\mathfrak{D}}_{n-i}(k)\otimes\widetilde{\Gb}_{i}(k)\delta_{j0}\right\},
\end{align}
with $\boldsymbol{\mathfrak{D}}_{n}(p,k)$ a c.c.~space matrix and $\boldsymbol{\mathfrak{D}}_{n}(k)$ a c.c.~space vector both given in Appendix~\ref{app:QExpansion}.

Summing the contribution from all LO diagrams the $Q^{2}$ part of the ``generic" LO form factor is given by

\begin{equation}
\frac{1}{2}\frac{\partial^{2}}{\partial Q^{2}}F_{0}(Q^{2})\Big{|}_{Q^{2}=0}=\frac{1}{2}\frac{\partial^{2}}{\partial Q^{2}}\left(F_{0}^{(a)}(Q^{2})+F_{0}^{(b)}(Q^{2})+F_{0}^{(c)}(Q^{2})\right)\Big{|}_{Q^{2}=0}.
\end{equation}
The NLO correction to the $Q^{2}$ part of the ``generic" form factor is
\begin{align}
&\frac{1}{2}\frac{\partial^{2}}{\partial Q^{2}}F_{1}(Q^{2})\Big{|}_{Q^{2}=0}=\frac{1}{2}\frac{\partial^{2}}{\partial Q^{2}}\left(F_{1}^{(a)}(Q^{2})+F_{1}^{(b)}(Q^{2})+F_{1}^{(c)}(Q^{2})+F_{1}^{(d)}(Q^{2})\right)\Big{|}_{Q^{2}=0}\\\nonumber
&\hspace{9cm}-\frac{\Sigma_{1}'(B)}{\Sigma_{0}'(B)}\frac{1}{2}\frac{\partial^{2}}{\partial Q^{2}}F_{0}(Q^{2})\Big{|}_{Q^{2}=0},
\end{align}
where the NLO diagrams are summed together and the LO contribution is multiplied by the NLO three-nucleon wavefunction renormalization.  Finally, including all $\nnlo$ contributions and multiplying the NLO term by the NLO three-nucleon wavefunction renormalization and the LO contribution by the $\nnlo$ three-nucleon wavefunction renormalization gives
\begin{align}
&\frac{1}{2}\frac{\partial^{2}}{\partial Q^{2}}F_{2}(Q^{2})\Big{|}_{Q^{2}=0}=\frac{1}{2}\frac{\partial^{2}}{\partial Q^{2}}\left(F_{2}^{(a)}(Q^{2})+F_{2}^{(b)}(Q^{2})+F_{2}^{(c)}(Q^{2})+F_{2}^{(d)}(Q^{2})\right)\Big{|}_{Q^{2}=0}\\\nonumber
&\hspace{2cm}-\frac{\Sigma_{1}'(B)}{\Sigma_{0}'(B)}\frac{1}{2}\frac{\partial^{2}}{\partial Q^{2}}F_{1}(Q^{2})\Big{|}_{Q^{2}=0}-\left(\frac{\Sigma_{2}'(B)}{\Sigma_{0}'(B)}+\frac{4}{3}M_{N}\widehat{H}_{2}\frac{\Sigma_{0}^{2}(B)}{\Sigma_{0}'(B)}\right)\frac{1}{2}\frac{\partial^{2}}{\partial Q^{2}}F_{0}(Q^{2})\Big{|}_{Q^{2}=0},
\end{align}
for the $\nnlo$ correction to the $Q^{2}$ part of the ``generic" form factor.

\subsection{Wigner-Symmetry: Consequences}

Additional information can be gleaned by going to the Wigner basis which is defined by
\begin{equation}
\Gamma_{n}^{(-)}(q)=\Gamma_{n,\psi\to Nt}(q)-\Gamma_{n,\psi\to Ns}(q)\quad,\quad\Gamma_{n}^{(+)}(q)=\Gamma_{n,\psi\to Nt}(q)+\Gamma_{n,\psi\to Ns}(q).
\end{equation}
At LO in the Wigner-SU(4) limit ($\gamma_{t}=\gamma_{s}$)~\cite{Bedaque:1999ve,Mehen:1999qs,Vanasse:2016umz} the component $\Gamma_{n}^{(+)}(q)=0$ and the LO triton charge form factor only depends on $\Gamma_{n}^{(-)}(q)$ giving the condition\footnote{Since the spin-singlet dibaryon is unphysical the $\Gamma_{n,\psi\to Ns}(q)$ vertex function can take an arbitrary phase.  Thus for some authors the roles of $\Gamma_{n}^{(+)}(q)$ and $\Gamma_{n}^{(-)}(q)$ are switched from the conventions of this work.}
\begin{align}
\label{eq:F0tritonWigner}
1=2\pi M_{N}\Gamma_{0}^{(-)}(q)\otimes\left\{\frac{\pi}{2}\frac{\delta(q-\ell)}{q^{2}\sqrt{\frac{3}{4}q^{2}-M_{N}B_{0}}}
-\frac{4}{q^{2}\ell^{2}-(q^{2}+\ell^{2}-M_{N}B_{0})^{2}}
\right\}\otimes\Gamma_{0}^{(-)}(\ell),
\end{align}
from Eq.~(\ref{eq:F0triton}).  Using this relationship and going to the Wigner-SU(4) limit for $\jjvH$ 	and $\jjvHe$ magnetic form factors gives the exact identities 
\begin{equation}
\label{eq:WignerResult}
F_{M}^{\jjvHe}(0)=(\kappa_{0}-\kappa_{1})=\mu_{n}\quad,\quad F_{M}^{\jjvH}(0)=(\kappa_{0}+\kappa_{1})=\mu_{p},
\end{equation}
for the LO three-nucleon magnetic form factors at $Q^{2}=0$.  In this work the magnetic form factors are normalized such that they give the three-nucleon magnetic moments in nuclear magnetons at $Q^{2}=0$.  The three-body wavefunction is spatially symmetric under the change of spatial coordinates of any two nucleons in the Wigner-SU(4) limit and nucleons of the same isospin state have opposite spin: as a result their magnetic moments cancel leaving the three-nucleon magnetic moment to be solely determined by the remaining unpaired nucleon, which is also known as the Schmidt-limit~\cite{Schmidt1937}.  Thus deviation from the Wigner-SU(4) limit is a measure of the ``asymmetry" of the spatial wavefunction, isospin breaking, and higher order two- and three-body currents.  Once Wigner-SU(4) symmetry is broken a small mixed symmetry $S'$-state is induced, which is not symmetric under the change of spatial coordinates of any two nucleons~\cite{Schiff:1964zz}.  The proton and neutron magnetic moments in nuclear magnetons are found to be $\mu_{p}=2.793$ and $\mu_{n}=-1.913$ respectively, while the three-nucleon magnetic moments are $\mu_{\jjvH}=2.979$ and $\mu_{\jjvHe}=-2.127$.  Thus $\mu_{p}$ is about 7\% away from $\mu_{\jjvH}$ and $\mu_{n}$ is about 11\% away from $\mu_{\jjvHe}$.  This implies that the Wigner-SU(4) limit is a good starting point to describe the three-nucleon system and a dual expansion in a Wigner-SU(4) symmetry breaking parameter and \EFT should yield order-by-order improvements~\cite{Vanasse:2016umz}.  A simple proof that Eq.~(\ref{eq:WignerResult}) follows as a consequence of Wigner-SU(4) symmetry is given in Appendix~\ref{app:Wigner}.

Going to the Wigner-SU(4) limit the NLO $\jjvH$ and $\jjvHe$ charge form factor at $Q^{2}=0$, Eq.~(\ref{eq:NLOform}), gives
\begin{align}
\label{eq:F0NLOtritonWigner}
0&=2\pi M_{N}\Gamma_{1}^{(-)}(q)\otimes\left\{\frac{\pi}{2}\frac{\delta(q-\ell)}{q^{2}\sqrt{\frac{3}{4}q^{2}-M_{N}B_{0}}}
-\frac{4}{q^{2}\ell^{2}-(q^{2}+\ell^{2}-M_{N}B_{0})^{2}}
\right\}\otimes\Gamma_{0}^{(-)}(\ell)\\\nonumber
&+2\pi M_{N}\Gamma_{0}^{(-)}(q)\otimes\left\{\frac{\pi}{2}\frac{\delta(q-\ell)}{q^{2}\sqrt{\frac{3}{4}q^{2}-M_{N}B_{0}}}
-\frac{4}{q^{2}\ell^{2}-(q^{2}+\ell^{2}-M_{N}B_{0})^{2}}
\right\}\otimes\Gamma_{1}^{(-)}(\ell)\\\nonumber
&-4\pi c_{W}\Gamma^{(-)}_{0}(q)\otimes\left\{\frac{\pi}{2}\frac{\delta(q-\ell)}{q^{2}}
\right\}\otimes\Gamma^{(-)}_{0}(\ell),
\end{align}
where in addition to the limit $\gamma_{t}=\gamma_{s}$ the limit $c_{W}=c_{0t}^{(0)}=c_{0s}^{(0)}$ is taken.  Using Eq.~(\ref{eq:F0NLOtritonWigner}) the NLO correction at $Q^{2}=0$ to the $\jjvH$ magnetic form factor in the Wigner-SU(4) limit is
\begin{equation}
F_{M,1}^{\jjvH}(0)=4\pi M_{N}\left(\frac{2}{3}\frac{c_{W}}{M_{N}}(\kappa_{0}+\kappa_{1})+\frac{1}{3}L_{2}+\frac{1}{3}L_{1}\right)\Gamma^{(-)}_{0}(q)\otimes\left\{\frac{\pi}{2}\frac{\delta(q-\ell)}{q^{2}}
\right\}\otimes\Gamma^{(-)}_{0}(\ell),
\end{equation}
and for the $\jjvHe$ magnetic form factor is
\begin{equation}
F_{M,1}^{\jjvHe}(0)=4\pi M_{N}\left(\frac{2}{3}\frac{c_{W}}{M_{N}}(\kappa_{0}-\kappa_{1})+\frac{1}{3}L_{2}-\frac{1}{3}L_{1}\right)\Gamma^{(-)}_{0}(q)\otimes\left\{\frac{\pi}{2}\frac{\delta(q-\ell)}{q^{2}}
\right\}\otimes\Gamma^{(-)}_{0}(\ell).
\end{equation}
Thus the NLO correction to the magnetic form factors in the Wigner-SU(4) limit at $Q^{2}=0$ can be entirely rewritten in terms of LO three-nucleon vertex functions.


\section{\label{sec:results}Results}
\subsection{Fitting $L_{1}$ and $L_{2}$}

To calculate the three-nucleon magnetic moments to NLO the LEC $L_{1}$ ($L_{2}$) of the isovector (isoscalar) two-body magnetic current term in Eq.~(\ref{eq:2BLagMag}) must be determined.  Typically, $L_{1}$ is fit to the cold $np$ capture cross-section ($\sigma_{np}$), which near threshold is given by~\cite{Chen:1999tn,Vanasse:2014sva}
\begin{align}
\label{eq:cross}
&\sigma_{np}=\frac{2\alpha\gamma_{t}^{6}}{|\vect{v}_{rel}|M_{N}^{3}}\left[\vphantom{\frac{1}{M_{N}}}|Y_{\mathrm{LO}}|^{2}+2\mathrm{Re}[Y_{\mathrm{LO}}^{*}Y_{\mathrm{NLO}}]\right],
\end{align}
where $Y_{\mathrm{LO}}$ ($Y_{\mathrm{NLO}}$) is the LO (NLO correction to the) isovector magnetic dipole moment, and $\vect{v}_{rel}$ is the relative velocity between the neutron and proton.  $Y_{\mathrm{LO}}$ at threshold in the $Z$-parametrization is given by~\cite{Vanasse:2013sda}
\begin{align}
Y_{\mathrm{LO}}=&\frac{2\kappa_{1}}{M_{N}\gamma_{t}^{2}}\sqrt{\gamma_{t}\pi}\left(1-\frac{\gamma_{t}}{\gamma_{s}}\right),
\end{align}
and its NLO correction depending on $L_{1}$ by 
\begin{align}
Y_{\mathrm{NLO}}=&\frac{2\kappa_{1}}{M_{N}\gamma_{t}^{2}}\sqrt{\gamma_{t}\pi}\frac{1}{2}\left((Z_{t}-1)-\frac{\gamma_{t}}{\gamma_{s}}\left[(Z_{t}-1)+(Z_{s}-1)\right]\right) \\\nonumber
&-\frac{L_{1}}{M_{N}\gamma_{s}}\sqrt{\gamma_{t}\pi}.
\end{align}
Ensuring reproduction of the experimental cold $np$ capture cross-section of $\sigma_{np}=334.2(5)$~mb~\cite{Cox:1965} at a neutron velocity of $v_{rel}=2200$~m/s yields $L_{1}=-6.90$~fm.

The value for $L_{2}$ is typically fit to the deuteron magnetic moment~\cite{Chen:1999tn} which to NLO in the $Z$-parametrization  in units of nuclear magnetons is given by~\cite{Vanasse:2013sda}\footnote{For similar expressions using different treatments of the LECs consult Refs.~\cite{Chen:1999tn,Beane:2000fi}}
\begin{equation}
\mu_{d}=\left(2Z_{t}\kappa_{0}+2L_{2}\gamma_{t}\right).
\end{equation}
Fitting to the experimental deuteron magnetic moment $\mu_{d}=0.85741 \frac{e}{2M_{N}}$ yields $L_{2}=-1.36$~fm.  By naturalness arguments the size of $L_{2}$ and $L_{1}$ should be approximately $1/m_{\pi}\sim 1.4$~fm.  Although $L_{2}$ is rather close to this value $L_{1}$ is significantly bigger.  However, since $L_{1}$ ($L_{2}$) is driven by $\kappa_{1}$ ($\kappa_{0}$) it should be divided by this scale.  Dividing $L_{1}$  by $2\kappa_{1}$ and $L_{2}$ by $2\kappa_{0}$ gives the values  -1.47~fm and -1.55~fm respectively, which are in line with naturalness expectations.

\subsection{Charge Radii of Three-Nucleon Systems}

The triton point charge radius has been calculated previously in \EFT~\cite{Platter:2005sj,Kirscher:2009aj,Vanasse:2016umz} up to $\nnlo$~\cite{Vanasse:2015fph}.  Here the results of Ref.~\cite{Vanasse:2015fph} are reviewed and the $\jjvHe$ point charge radius to $\nnlo$ in the absence of Coulomb is given.  The point charge radius squared is given by
\begin{equation}
\left<\delta r_{C}^{2}\right>^{{}^{A}Z}_{n}=-\frac{6}{Z^{^AZ}}\left(\frac{1}{2}\frac{\partial^{2}}{\partial Q^{2}}F_{C,n}^{^{A}Z}(Q^{2})\Big{|}_{Q=0}\right),
\end{equation}
where $Z^{^AZ}$ is the number of protons in the nucleus, $n=0$ is the LO term, $n=1$ is the NLO correction, and $n=2$ is the $\nnlo$ correction.  Taking the square root yields the three-nucleon point charge radius, which up to to $\nnlo$ is 
\begin{equation}
\delta r_{C}^{^AZ}=\sqrt{\left<\delta r_{C}^{2}\right>^{^AZ}_{0}}\left(\underbrace{\vphantom{\frac{1}{2}\frac{\left<r_{C}^{2}\right>^{^AZ}_{1}}{\left<\delta r_{C}^{2}\right>^{^AZ}_{0}}}1}_{\mathrm{LO}}+\underbrace{\frac{1}{2}\frac{\left<\delta r_{C}^{2}\right>^{^AZ}_{1}}{\left<\delta r_{C}^{2}\right>^{^AZ}_{0}}}_{\mathrm{NLO}}+\underbrace{\frac{1}{2}\frac{\left<\delta r_{C}^{2}\right>^{^AZ}_{2}}{\left<\delta r_{C}^{2}\right>^{^AZ}_{0}}-\frac{1}{8}\left(\frac{\left<\delta r_{C}^{2}\right>^{^AZ}_{1}}{\left<\delta r_{C}^{2}\right>^{^AZ}_{0}}\right)^{2}}_{\nnlo}+\cdots\right).
\end{equation}

The three-nucleon charge radius $r_{C}^{^AZ}$ is related to the three-nucleon point charge radius $\delta r_{C}^{^AZ}$ by
\begin{equation}
\label{eq:pointcharge}
\left<\left(\delta r_{C}^{{}^{A}Z}\right)^{2}\right>=\left<\left(r_{C}^{^{A}Z}\right)^{2}\right>-Z^{^AZ}\left<r_{p}^{2}\right>-N^{^AZ}\left<r_{n}^{2}\right>,
\end{equation}
where $N^{^AZ}$ is the number of neutrons in the nucleus, $r_{p}=0.8783\pm0.0086$~fm~\cite{Angeli201369} is the proton charge radius, $r_{n}^{2}=-0.1149\pm0.0027$~fm$^{2}$~\cite{Angeli201369} is the neutron charge radius squared, and $r_{C}^{^AZ}$ is the three-nucleon charge radius.  For $\jjvH$ ($\jjvHe$) the experimental charge radius is $r_{C}^{\jjvH}=1.7591(363)$~fm ($r_{C}^{\jjvHe}=1.9961(30)$~fm)~\cite{Angeli201369} and the resulting point charge radius is $\delta r_{C}^{\jjvH}=1.5978(40)$~fm ($\delta r_{C}^{\jjvHe}=1.77527(540)$~fm).  The point charge radius of $\jjvH$ and $\jjvHe$ up to $\nnlo$ are compared with experiment in Table~\ref{tab:chargeradius}. 
\begin{table}[hbt]
\begin{tabular}{|c|cc|}\hline
& $\delta r_{C}^{\jjvH}$~fm & $\delta r_{C}^{\jjvHe}$~fm\\\hline
LO & 1.14(20) & 1.26(22) \\
NLO & 1.59(9) & 1.72(10) \\
$\nnlo$ & 1.62(4)  & 1.74(4)\\\hline
Exp & 1.5978(40) & 1.77527(540) \\\hline
\end{tabular}
\caption{\label{tab:chargeradius}$\jjvH$ and $\jjvHe$ point charge radius up to $\nnlo$ compared to experimental data~\cite{Angeli201369}.}
\end{table}
Overlap within errors between theory and experiment is seen at NLO and $\nnlo$ for the $\jjvH$ and $\jjvHe$ point charge radius.  The LO point charge radius for $\jjvH$ and $\jjvHe$ under-predict the experimental values.  However, as noted in Ref.~\cite{Vanasse:2015fph} the correct value for the LO point charge radius is obtained in the unitary limit.  Thus, even though the LO values seem too small they are reasonable.  Despite the absence of Coulomb interactions in these calculations good agreement is found between theory and experiment for the $\jjvHe$ point charge radius.  Errors due to excluding Coulomb corrections are roughly of the size $\alpha M_{n}/(2p^{*}_{\jjvHe})\sim 4\%$, where $p^{*}_{\jjvHe}=\sqrt{M_{N}E_{\jjvHe}}$ is the binding momentum of $\jjvHe$, for the $\jjvHe$ binding energy $E_{\jjvHe}=7.718$~MeV~\cite{Wapstra:1985zz}, and the factor of two comes from taking the square root of the amplitude for the charge radius.

The error estimate for amplitudes  in \EFT follows from the expansion parameter $\frac{1}{2}(Z_{t}-1)=0.345$ leading to a 35\% error estimate at LO, a 12\% error estimate at NLO, and a 4\% error estimate at $\nnlo$.\footnote{Note, this error estimate is more conservative than that used in Ref.~\cite{Vanasse:2015fph}, which calculated the $\jjvH$ charge radius to $\nnlo$.}  Error estimates for charge and magnetic radii are half of these values since it is given by the square root of an amplitude.  Likewise, the error estimate for cross-sections is doubled since it is obtained from squaring an amplitude.  The error for the $\nnlo$ three-nucleon charge radii comes from a $\sim\!\!1\%$ error from cutoff variation and a $\sim\!\!2\%$ error from the \EFT expansion.  This slight cutoff variation is due to either a slow divergence or convergence of the $\nnlo$ three-nucleon point charge radii.  To answer this conclusively either a detailed asymptotic analysis must be performed or a calculation to higher cutoffs.  However, for cutoffs $\Lambda > 10^{6}$~MeV numerical issues are currently encountered and reliable calculations cannot be performed.  To go to higher cutoffs new numerical techniques will be required.  Finally, the $\jjvH$ charge radius has been calculated previously in \EFT yielding the LO prediction of $2.1\pm0.6$~fm~\cite{Platter:2005sj} and the NLO prediction $1.6\pm0.2$~fm~\cite{Kirscher:2009aj}.  The former result used wavefunction techniques and the latter was a position space calculation.  A comparison of various techniques for the $\jjvH$ point charge radius can be found in Ref.~\cite{Vanasse:2015fph}.

\subsection{Observables from The Magnetic Form Factor}

The three-nucleon magnetic form factor given in Eq.~(\ref{eq:generic}) when $X=M$ can be expanded perturbatively as
\begin{equation}
F_{M}^{^{A}Z}(Q^{2})=F_{M,0}^{^{A}Z}(Q^{2})+F_{M,1}^{^{A}Z}(Q^{2})+\cdots,
\end{equation}
the three-nucleon magnetic moment can be expanded perturbatively giving
\begin{equation}
\mu_{^AZ}=\mu^{^AZ}_{0}+\mu^{^AZ}_{1}+\cdots,
\end{equation}
and the three-nucleon point magnetic radius squared can be expanded perturbatively yielding
\begin{equation}
\left<\delta r_{M}^{2}\right>^{{}^{A}Z}=\left<\delta r_{M}^{2}\right>^{{}^{A}Z}_{0}+\left<\delta r_{M}^{2}\right>^{{}^{A}Z}_{1}+\cdots,
\end{equation}
where the terms with subscript ``0" (``1") are the LO contribution (NLO correction).  Using this perturbative expansion the LO three-nucleon magnetic moment is
\begin{equation}
\mu^{^AZ}_{0}=F_{M,0}^{^{A}Z}(0),
\end{equation}
and it NLO correction
\begin{equation}
\mu^{^AZ}_{1}=F_{M,1}^{^{A}Z}(0).
\end{equation}
The LO three-nucleon point magnetic radius squared is given by
\begin{equation}
\left<\delta r_{M}^{2}\right>^{{}^{A}Z}_{0}=-\frac{6}{\mu^{^AZ}_{0}}\left(\frac{1}{2}\frac{\partial^{2}}{\partial Q^{2}}F_{M,0}^{^{A}Z}(Q^{2})\Big{|}_{Q=0}\right),
\end{equation}
and its NLO correction by
\begin{equation}
\left<\delta r_{M}^{2}\right>^{{}^{A}Z}_{1}=-\frac{6}{\mu^{^AZ}_{0}}\left(\frac{1}{2}\frac{\partial^{2}}{\partial Q^{2}}F_{M,1}^{^{A}Z}(Q^{2})\Big{|}_{Q=0}\right)-\frac{\mu^{^AZ}_{1}}{\mu^{^AZ}_{0}}\left<\delta r_{M}^{2}\right>^{^AZ}_{0}.
\end{equation}
Finally, the resulting NLO three-nucleon point magnetic radius is given by
\begin{equation}
\delta r_{M}^{^AZ}=\sqrt{\left<\delta r_{M}^{2}\right>^{{}^{A}Z}_{0}}\left(\underbrace{\vphantom{\frac{\left<\delta r_{M}^{2}\right>^{{}^{A}Z}_{1}}{\left<\delta r_{M}^{2}\right>^{{}^{A}Z}_{0}}}1}_{\mathrm{LO}}+\underbrace{\frac{1}{2}\frac{\left<\delta r_{M}^{2}\right>^{{}^{A}Z}_{1}}{\left<\delta r_{M}^{2}\right>^{{}^{A}Z}_{0}}}_{\mathrm{NLO}}+\cdots\right).
\end{equation}

From the $\jjvH$ and $\jjvHe$ magnetic moments the isoscalar and isovector magnetic moment can be defined as
\begin{equation}
\mu_{s}=\frac{1}{2}\left(\mu_{\jjvHe}+\mu_{\jjvH}\right),
\end{equation}
and
\begin{equation}
\mu_{v}=\frac{1}{2}\left(\mu_{\jjvHe}-\mu_{\jjvH}\right),
\end{equation}
respectively.  The isoscalar magnetic moment only depends on $L_{2}$ and $\kappa_{0}$ up to NLO, while the isovector magnetic moment only depends on $L_{1}$ and $\kappa_{1}$ up to NLO.  At LO the isoscalar magnetic moment in nuclear magnetons is given by $\mu_{s}=0.440(152)$ and fitting $L_{2}$ to the deuteron magnetic moment gives the NLO value $\mu_{s}= 0.421(50)$.  This agrees well with the experimental value of $\mu_{s}=0.426$.  $\mu_{v}$ is compared with experiment in Table~\ref{tab:isomoments}. 
\begin{table}[hbt]
\begin{tabular}{|c|cc|}\hline
 &  $\mu_{v}$ & $L_{1}$ fit\\\hline
LO &  -2.31(80) & N/A \\ 
NLO &  -2.20(26) & $\sigma_{np}$ \\
NLO &  -2.56(31) & $\mu_{\jjvH}$ \\
NLO &  -2.50(30) & $\sigma_{np}$ and $\mu_{\jjvH}$\\\hline
Exp &  -2.55 & N/A\\\hline
\end{tabular}
\caption{\label{tab:isomoments}Table of three-nucleon isovector magnetic moments compared to experiment.  The first NLO row is for $L_{1}$ fit to $\sigma_{np}$, the second NLO row is for $L_{1}$ fit to $\mu_{\jjvH}$, and the final NLO row is $L_{1}$ fit to both $\sigma_{np}$ and $\mu_{\jjvH}$.}
\end{table}
The first NLO row is for $L_{1}$ fit to $\sigma_{np}$ and it is observed that $\mu_{v}$ is slightly under-predicted at NLO.  In the second NLO row $L_{1}=-5.62$~fm is fit to the experimental $\jjvH$ magnetic moment.  The last NLO row is given by $L_{1}=-5.83$~fm which is a best fit to both $\sigma_{np}$ and $\mu_{\jjvH}$, where the relative error for $\sigma_{np}$ and $\mu_{\jjvH}$ between theory and experiment is minimized.  For both of these choices of $L_{1}$ agreement is found between theory at NLO and experiment for $\mu_{v}$.  The value for $L_{1}$ fit to $\sigma_{np}$ and $\mu_{\jjvH}$ simultaneously with its associated \EFT error is $L_{1}=-5.83\pm2.01$~fm.  Values for $L_{1}$ fit to just $\sigma_{np}$ or $\mu_{\jjvH}$ are encompassed within this error.

\subsection{Magnetic Moments and Radii of Three-Nucleon Systems}

The LO and NLO calculation of the three-nucleon magnetic radii in \EFT treat nucleons as point particles and hence the resulting values are called the point magnetic radii.  Contributions from the nucleon magnetic radii occur at $\nnlo$ in \EFT and are given by the Lagrangian
\begin{equation}
\mathcal{L}^{mag}_{1,2}=-\frac{1}{6}\left<r_{M}^{2}\right>_{p}\mu_{p}\hat{N}^{\dagger}\left(\frac{1+\tau_{3}}{2}\right)\vec{\sigma}\cdot\vec{\nabla}^{2}\mathbf{B}\hat{N}+-\frac{1}{6}\left<r_{M}^{2}\right>_{n}\mu_{n}\hat{N}^{\dagger}\left(\frac{1-\tau_{3}}{2}\right)\vec{\sigma}\cdot\vec{\nabla}^{2}\mathbf{B}\hat{N},
\end{equation}
where $\sqrt{\left<r_{M}^{2}\right>_{p}}=0.776(34)(17)$~fm~\cite{Olive:2016xmw,Lee:2015jqa} is the proton magnetic radius and  $\sqrt{\left<r_{M}^{2}\right>_{n}}=0.864^{+0.009}_{-0.008}$~fm~\cite{Olive:2016xmw,Epstein:2014zua} is the neutron magnetic radius.  Using this interaction the full magnetic radius of $\jjvH$ is
\begin{equation}
\label{eq:3Hmagrad}
\left<r_{M}^{2}\right>^{\jjvH}=\left<\delta r_{M}^{2}\right>^{\jjvH}+\frac{1}{2}\frac{\mu_{p}}{\mu_{\jjvH}}\left(\frac{\mu_{s}^{\mathrm{LO}}}{\kappa_{0}}-\frac{\mu_{v}^{\mathrm{LO}}}{\kappa_{1}}\right)\left<r_{M}^{2}\right>_{p}+\frac{1}{2}\frac{\mu_{n}}{\mu_{\jjvH}}\left(\frac{\mu_{s}^{\mathrm{LO}}}{\kappa_{0}}+\frac{\mu_{v}^{\mathrm{LO}}}{\kappa_{1}}\right)\left<r_{M}^{2}\right>_{n},
\end{equation}
and for $\jjvHe$ is
\begin{equation}
\label{eq:3Hemagrad}
\left<r_{M}^{2}\right>^{\jjvHe}=\left<\delta r_{M}^{2}\right>^{\jjvHe}+\frac{1}{2}\frac{\mu_{p}}{\mu_{\jjvHe}}\left(\frac{\mu_{s}^{\mathrm{LO}}}{\kappa_{0}}+\frac{\mu_{v}^{\mathrm{LO}}}{\kappa_{1}}\right)\left<r_{M}^{2}\right>_{p}+\frac{1}{2}\frac{\mu_{n}}{\mu_{\jjvHe}}\left(\frac{\mu_{s}^{\mathrm{LO}}}{\kappa_{0}}-\frac{\mu_{v}^{\mathrm{LO}}}{\kappa_{1}}\right)\left<r_{M}^{2}\right>_{n},
\end{equation}
where $\mu_{s}^{\mathrm{LO}}$ ($\mu_{v}^{\mathrm{LO}}$) is the LO three-nucleon isoscalar (isovector) magnetic moment in \EFT.  From Eq.~(\ref{eq:WignerResult}) it can be shown that in the Wigner-SU(4) limit these equations reduce to
\begin{equation}
\left<r_{M}^{2}\right>^{\jjvH}=\left<\delta r_{M}^{2}\right>^{\jjvH}+\left<r_{M}^{2}\right>_{p},
\end{equation}
and
\begin{equation}
\left<r_{M}^{2}\right>^{\jjvHe}=\left<\delta r_{M}^{2}\right>^{\jjvHe}+\left<r_{M}^{2}\right>_{n}.
\end{equation}
This result is analogous to the magnetic moments in the Wigner-SU(4) limit since the correction to the three-nucleon point magnetic radius only depends on the magnetic radius of the unpaired nucleon.  Although contributions from the nucleon magnetic radii are strictly $\nnlo$ we include them at LO and NLO to compare with experimental results that include such nucleon structure.  For the three-nucleon charge radii the experimental nucleon charge radii are subtracted from the experimental three-nucleon charge radii to give the ``experimental" three-nucleon point charge radii (See Eq.~(\ref{eq:pointcharge})) that are then compared with theory.  However, unlike Eq.~(\ref{eq:pointcharge}) Eqs.~(\ref{eq:3Hmagrad}) and (\ref{eq:3Hemagrad}) depend on theoretical numbers.  Therefore, the nucleon magnetic radii are added to the theoretical point magnetic radii to get the full magnetic radius rather than subtracting them from the experimental three-nucleon magnetic radii and pollute experimental numbers with theoretical numbers.  

Due to gauge invariance the three-nucleon charge is reproduced exactly at LO and higher order one-, two- and, three-body currents are arranged such that the three-nucleon charge remains fixed, or simply put $F_{C,n}^{^{A}Z}(Q^{2}=0)=0$ for $n\geq 1$.  However, no such condition exists for the magnetic form factor, thus the three nucleon magnetic moments receive non-zero contributions from higher order one-,two-, and three-body currents and they are not reproduced exactly at LO.  This is why Eq.~(\ref{eq:pointcharge}) contains no theoretical numbers, whereas Eqs.~(\ref{eq:3Hmagrad}) and (\ref{eq:3Hemagrad}) do.

The values of the three-nucleon magnetic moments, three-nucleon magnetic radii, and $\sigma_{np}$ up to NLO are given in Table~\ref{tab:magresults}.
%
%
\begin{table}[hbt]
\begin{tabular}{|c|cccccc|}\hline
 & $\mu_{\jjvH}$ & $\mu_{\jjvHe}$ & $r_{M}^{\jjvH}$~fm & $r_{M}^{\jjvHe}$~fm & $\sigma(np\to d\gamma)$~mb & $L_{1}$ fit \\\hline
 LO & 2.75(95) & -1.87(73) & 1.40(24) & 1.49(26) & $325.2\pm 225.6$ & N/A\\
 NLO & 2.62(31) & -1.78(21) & 1.83(11) & 1.92(11) & $334.2\pm 79.7$ & $\sigma_{np}$\\
 NLO & 2.98(36) & -2.14(25) & 1.77(11) & 1.83(11) & $370.47 \pm 88.4$ & $\mu_{\jjvH}$\\  
 NLO & 2.92(35) & -2.08(25) & 1.78(11) & 1.85(11) & $364.5 \pm 87.0$ & $\sigma_{np}$ and $\mu_{\jjvH}$\\\hline
 Exp & 2.979 & -2.127 & 1.840(182)~\cite{Sick:2001rh} & 1.965(154)~\cite{Sick:2001rh} & 334.2(5)~\cite{Cox:1965} & N/A\\\hline
\end{tabular}
\caption{\label{tab:magresults}Values of three-nucleon magnetic moments, three nucleon magnetic radii, and $\sigma_{np}$ to NLO compared to experiment.  The different NLO rows are different fits for $L_{1}$ as described in Table~\ref{tab:isomoments}. }
\end{table}
There are three NLO rows, arranged as in Table~\ref{tab:isomoments}, corresponding to different methods for fitting the $L_{1}$ coefficient.  For all of these rows $L_{2}$ is fit to the deuteron magnetic moment.  Fitting $L_{1}$ to $\sigma_{np}$ it is observed that the three-nucleon magnetic moments are slightly under-predicted at NLO, while fitting $L_{1}$ to $\mu_{\jjvH}$ or $\mu_{\jjvH}$ and $\sigma_{np}$ leads to agreement between the three-nucleon magnetic moments and experiment.  For all choices of $L_{1}$ the three-nucleon magnetic radii at NLO overlap within errors with the experimental values, in part due to their relatively large experimental error.  All NLO results for the magnetic radii have a smaller estimated theoretical error than the current experimental error.  The results for $L_{1}$ fit to $\mu_{\jjvH}$ and $\sigma_{np}$ are taken as the predictions of \EFT at NLO in this work.


\section{\label{sec:conclusions} Conclusions}
In this work it was demonstrated how the zeroth (\emph{i.e.} charge of charge form factor and magnetic moment of magnetic form factor)  and second (\emph{i.e.} charge and magnetic radius of charge and magnetic form factor respectively)  moment of a generic form factor for a three-body system coming from an external current via a non-derivative coupling can be calculated in srEFT. This was carried out for three-nucleon systems in \EFT.  Extension of this work to calculate the generic form factor for arbitrary $Q^{2}$ values is straightforward using the work of Refs.~\cite{Hagen:2013xga,Vanasse:2015fph,Vanasse:2016hgn} but of limited interest in \EFT which in only valid for $Q\lesssim 0.7~\mathrm{fm}^{-1}$.  In addition by calculating the zeroth and second moments directly the number of integrals is reduced and results at larger cutoffs can be calculated without numerical issues.  Using the methods of this work the point charge radii of the three-nucleon system was calculated to $\nnlo$, while the magnetic moments and magnetic radii were calculated to NLO.

The point charge radius of $\jjvH$ ($\jjvHe$) is 1.14(20)~fm (1.26(22)~fm) at LO, 1.59(9)~fm (1.72(4)~fm) at NLO, and 1.62(4)~fm (1.74(4)~fm) at $\nnlo$.\footnote{The point charge radius of $\jjvH$ was calculated previously in Ref.~\cite{Vanasse:2015fph} but is shown here with the more conservative error estimates of this work.}   NLO and $\nnlo$ values both agree within errors with the experimental point charge radius of 1.5978(40)~fm (1.7763(54)~fm) for $\jjvH$ ($\jjvHe$).  In this work Coulomb interactions were not included in the calculation of $\jjvHe$ properties.  Coulomb corrections should be roughly a 4\% correction and can be included perturbatively~\cite{Konig:2015aka,Konig:2016iny}.  The error estimate at $\nnlo$ is due to a small amount of observed cutoff variation ($\sim$1\%) and \EFT error ($\sim$2\%).  Cutoff variation at $\nnlo$ is either a consequence of slow convergence or divergence.  To answer the question of which a detailed asymptotic analysis or a calculation to higher cutoffs needs to be carried out.  Definitively answering this issue at $\nnlo$ is relegated to future work.

The magnetic moment of $\jjvH$ ($\jjvHe$) at LO is $2.75(95)$ ($-1.87(73)$) in nuclear magnetons.  This agrees with the experimental magnetic moment for  $\jjvH$ ($\jjvHe$)  of 2.979 (-2.127).  The LO magnetic radius of $\jjvH$ ($\jjvHe$) is 1.40(24)~fm (1.49(26)~fm), which under-predicts the experimental value of 1.840(182)~fm (1.965(154)~fm)~\cite{Sick:2001rh}.  This under-prediction is in line with what is observed for the three-nucleon point charge radius at LO.  Indeed a large jump is observed in the LO to NLO three-nucleon magnetic radius just as for the three-nucleon point charge radius.  At NLO the isoscalar two-body current $L_{2}$ is fit to the deuteron magnetic moment.  Fitting the isovector two-body current $L_{1}$ to $\mu_{\jjvH}$ and $\sigma_{np}$ ($L_{1}=-5.83\pm2.01$~fm) yields the NLO $\jjvH$ ($\jjvHe$) magnetic moment 2.92(35) (-2.08(25)), the NLO cold $np$ capture cross-section $364.5\pm 87.0$~mb, and the NLO $\jjvH$ ($\jjvHe$) magnetic radius 1.78(11)~fm (1.85(11)~fm), which are all consistent with experiment. 

Using coordinate space techniques the three-nucleon magnetic moments have been calculated previously by Kirscher \emph{et al.}~\cite{Kirscher:2017fqc} in \EFT to NLO for different pion masses by fitting to Lattice QCD data.  Calculating at the physical pion mass Kirscher \emph{et al.} fit $L_{1}$ to $\mu_{\jjvH}$ and found good predictions for the $\jjvHe$ magnetic moment and the $\jjvH$ and $\jjvHe$ magnetic polarizabilities at NLO in which Coulomb interactions were included.  They also considered the three-nucleon charge radii at NLO and found good agreement with experiment.  However, they did not consider the value of $\sigma_{np}$ with their choice of $L_{1}$.

Due to gauge symmetry the three-nucleon charge form factor at LO gives the three-nucleon charge at $Q^{2}$=0 and all higher order corrections give zero contribution at $Q^{2}$=0.  This gives exact analytical expressions involving the properly renormalized three-nucleon vertex functions.  Going to the Wigner-SU(4) limit these identities for the charge form factor can be used to make predictions about the magnetic form factor.  In particular it is found that the $\jjvH$ ($\jjvHe$) magnetic moment is the proton (neutron) magnetic moment in the Wigner-SU(4) limit reproducing the Schmidt-limit~\cite{Schmidt1937}.  Comparing this to the experimental values of three-nucleon magnetic moments provides added evidence that the Wigner-SU(4) limit is a good starting point to describe three-nucleon systems.  Corrections breaking Winger-SU(4) symmetry can be added perturbatively as in Ref.~\cite{Vanasse:2016umz}.  It is also demonstrated that the NLO correction to the magnetic moment can be written entirely in terms of LO quantities in the Wigner-SU(4) limit.  These results in the Wigner-symmetric limit should be taken as an essential benchmark since any technique that is able to calculate three-nucleon magnetic moments in the Wigner-SU(4) limit should reproduce them exactly up to to numerical error.

The most accurate determination to date of three-nucleon magnetic radii is from magnetic form factors measured from electron scattering~\cite{Amroun:1994qj}.  Extracting the magnetic form factor requires looking at the angular distribution of scattered electrons and subtracting out the much larger isotropic contribution from the charge form factor, which leads to the larger uncertainties for the magnetic radii vs.~the charge radii.  Future experiments involving the hyperfine splitting of atomic $S$-wave states in muonic $\jjvH$ ($\mu\jjvH^{+}$) and $\jjvHe$ ($\mu\jjvHe^{+}$) offer the possibility of more precise measurements of the three-nucleon magnetic radii~\cite{Karshenboim:2014vea,Antognini:2015vxo}.  The NLO \EFT prediction of the three-nucleon magnetic radii has an approximate 6\% error, which is already smaller than the experimental error for the magnetic radius of $\jjvH$ ($\jjvHe$) of $\sim\!\!10\%$ ($\sim\!\!8\%$).  \EFT offers the possibility of precision calculations with controlled errors for three-nucleon properties.  A $\nnlo$ \EFT calculation of three-nucleon magnetic radii including perturbative Coulomb corrections~\cite{Konig:2015aka} would give a result accurate to about 2\%.

Using spectroscopic data on $\mu\jjvHe^{+}$ the $\jjvHe$ charge radius can in principle be determined to hundred-fold better accuracy than from current electron scattering experiments~\cite{Franke:2017tpc}.  However, extraction from spectroscopic data is hampered by the relative uncertainty of two photon exchange (TPE) contributions.  The best current theoretical determinations of TPE are accurate to about 3\%~\cite{Carlson:2016cii,Dinur:2015vzv,Hernandez:2016jlh}.  A N$^{3}$LO \EFT calculation of TPE can improve on this accuracy by a factor of two.  Measurement of the $\jjvHe$ charge radius from $\mu\jjvHe^{+}$ will give insight into the so called ``proton radius puzzle"~\cite{Pohl:2013yb} in which a seven standard deviation discrepancy is seen between the proton charge radius determined from electron scattering and Hydrogen spectroscopy~\cite{Mohr:2012tt} vs. muonic-Hydrogen spectroscopy~\cite{Antognini:1900ns}.

\acknowledgments{I would like to thank Roxanne Springer, Doron Gazit, Hilla De-Leon, and Daniel Phillips for useful discussions during the course of this work.  In  addition I  would  like to thank Daniel Phillips for valuable comments on the manuscript.  I would also like to thank the ExtreMe Matter Institute EMMI at the GSI Helmholtz Centre for Heavy Ion Research and the Kalvi Institute for Theoretical Physics program, \emph{Frontiers in Nuclear Physics}, for support during the completion of this work.  This material is based upon work supported by the U.S. Department of
Energy, Office of Science, Office of Nuclear Physics, under Award Number DE-FG02-93ER40756}

\appendix

\section{\label{app:chi}}

To derive the values in Table~\ref{tab:LOvalues} the spin-isospin operator in c.c.~space for each diagram in Fig.~\ref{fig:FormFactorLO} must be projected onto the doublet $S$-wave channel.  The spin-isospin c.c.~space operator for diagram-(a) for minimally coupled $\hat{A}_{0}$ photons is
\begin{equation}
\left(\frac{1+\tau_{3}}{2}\right)^{a}_{b}\delta^{\alpha}_{\beta}\delta^{ij},
\end{equation}
and for magnetically coupled photons
\begin{equation}
\left(\kappa_{0}+\tau_{3}\kappa_{1}\right)^{a}_{b}\left(\sigma_{n}\right)^{\alpha}_{\beta}\delta^{ij},
\end{equation}
where $\alpha$ ($\beta$) is the initial (final) nucleon spin, $a$ ($b$) is the initial (final) nucleon isospin, and $i$ ($j$) is the initial (final) dibaryon polarization.  Projecting these operators into the doublet $S$-wave channel using the projectors in Ref.~\cite{Griesshammer:2004pe} yields
\begin{equation}
\frac{1}{3}\left(\begin{array}{cc}
\sigma_{j}  & 0 \\[-1.5 mm]
0 & \tau_{B} 
\end{array}\right)
\left(\begin{array}{cc}
\left(\frac{1+\tau_{3}}{2}\right)\delta_{ij} & 0 \\[0 mm] 
0 & \left(\frac{1+\tau_{3}}{2}\right)\delta_{AB}
\end{array}\right)
\left(\begin{array}{cc}
\sigma_{i}  & 0 \\[-1.5 mm]
0 & \tau_{A} 
\end{array}\right)=
\left(\begin{array}{cc}
\left(\frac{1+\tau_{3}}{2}\right)  & 0 \\[-1.5 mm]
0 & \frac{1}{3}\left(\frac{3-\tau_{3}}{2}\right)
\end{array}\right),
\end{equation}
for minimally coupled $\hat{A}_{0}$ photons and
\begin{align}
&\frac{1}{3}\left(\begin{array}{cc}
\sigma_{j}  & 0 \\[-1.5 mm]
0 & \tau_{B} 
\end{array}\right)
\left(\begin{array}{cc}
(\kappa_{0}+\tau_{3}\kappa_{1})\sigma_{n}\delta_{ij} & 0 \\[0 mm] 
0 & (\kappa_{0}+\tau_{3}\kappa_{1})\sigma_{n}\delta_{AB}
\end{array}\right)
\left(\begin{array}{cc}
\sigma_{i}  & 0 \\[-1.5 mm]
0 & \tau_{A} 
\end{array}\right)=\\\nonumber
&\hspace{5cm}\left(\begin{array}{cc}
-\frac{1}{3}\left(\kappa_{0}+\tau_{3}\kappa_{1}\right)  & 0 \\[-1.5 mm]
0 & \frac{1}{3}\left(3\kappa_{0}-\tau_{3}\kappa_{1}\right)
\end{array}\right)\sigma_{n},
\end{align}
for magnetically coupled photons.  Choosing $\tau_{3}=1$ ($\tau_{3}=-1$) gives the coefficients $a_{11}$ and $a_{22}$ for $\jjvHe$ ($\jjvH$) in Table~\ref{tab:LOvalues}.  The Pauli matrix $\sigma_{n}$ couples to the magnetic field $\mathbf{B}_{n}$ not shown here.

The spin-isospin c.c.~space operator for diagram-(b) of Fig.~\ref{fig:FormFactorLO} for minimally coupled $\hat{A}_{0}$ photons is given by
\begin{equation}
\left[{P_{i}^{(w)}}^{\dagger}\left(\frac{1+\tau_{3}}{2}\right)P_{j}^{(x)}\right]^{\alpha a}_{\beta b},
\end{equation}
and for magnetically coupled photons by
\begin{equation}
\left[{P_{i}^{(w)}}^{\dagger}\left(\kappa_{0}+\tau_{3}\kappa_{1}\right)\sigma_{n}P_{j}^{(x)}\right]^{\alpha a}_{\beta b},
\end{equation}
where $P_{j}^{(x)}=\sqrt{8}P_{j}$ ($P_{j}^{(x)}=\sqrt{8}\bar{P}_{j}$) for $x=t$ ($x=s$) in the spin-triplet iso-singlet (spin-singlet iso-triplet) channel.  Here the indices ``$i$" and ``$j$" are either spinor or isospinor indices depending on the values of ($x$) and ($w$).  The values of ($x$) and ($w$) give the matrix element of the c.c.~space matrix.  Projecting onto the doublet $S$-wave channel gives
\begin{equation}
\frac{1}{3}\left(\begin{array}{cc}
\sigma_{j}  & 0 \\[-1.5 mm]
0 & \tau_{B} 
\end{array}\right)
\left(\begin{array}{cc}
\left(\frac{1-\tau_{3}}{2}\right)\sigma_{i}\sigma_{j} & \tau_{A}\left(\frac{1-\tau_{3}}{2}\right)\sigma_{j} \\[0 mm] 
\left(\frac{1-\tau_{3}}{2}\right)\tau_{B}\sigma_{i} & \tau_{A}\left(\frac{1-\tau_{3}}{2}\right)\tau_{B}
\end{array}\right)
\left(\begin{array}{cc}
\sigma_{i}  & 0 \\[-1.5 mm]
0 & \tau_{A} 
\end{array}\right)=
\left(\begin{array}{cc}
-\left(\frac{1-\tau_{3}}{2}\right)  & \left(\frac{3+\tau_{3}}{2}\right) \\[-1.5 mm]
\left(\frac{3+\tau_{3}}{2}\right) & -\frac{1}{3}\left(\frac{3+5\tau_{3}}{2}\right)
\end{array}\right),
\end{equation}
for minimally coupled $\hat{A}_{0}$ photons and
\begin{align}
&\frac{1}{3}\left(\begin{array}{cc}
\sigma_{j}  & 0 \\[-1.5 mm]
0 & \tau_{B} 
\end{array}\right)
\left(\begin{array}{cc}
-(\kappa_{0}-\tau_{3}\kappa_{1})\sigma_{i}\sigma_{n}\sigma_{j} & -\tau_{A}(\kappa_{0}-\tau_{3}\kappa_{1})\sigma_{n}\sigma_{j} \\[0 mm] 
-(\kappa_{0}-\tau_{3}\kappa_{1})\tau_{B}\sigma_{i}\sigma_{n} & -\tau_{A}(\kappa_{0}-\tau_{3}\kappa_{1})\tau_{B}\sigma_{n}
\end{array}\right)
\left(\begin{array}{cc}
\sigma_{i}  & 0 \\[-1.5 mm]
0 & \tau_{A} 
\end{array}\right)=\\\nonumber
&\left(\begin{array}{cc}
-\frac{5}{3}(\kappa_{0}-\tau_{3}\kappa_{1})  & \frac{1}{3}(3\kappa_{0}+\tau_{3}\kappa_{1}) \\[-1.5 mm]
\frac{1}{3}(3\kappa_{0}+\tau_{3}\kappa_{1}) & \frac{1}{3}(3\kappa_{0}+5\tau_{3}\kappa_{1})
\end{array}\right)\sigma_{n},
\end{align}
for magnetically coupled photons.

The spin isospin c.c.~space operator for diagram-(c) of Fig.~\ref{fig:FormFactorLO} for minimally coupled $\hat{A}_{0}$ photons is given by
\begin{equation}
\frac{1}{2}\mathrm{Tr}\left[P_{j}^{(x)}\left(\frac{1+\tau_{3}}{2}\right){P_{i}^{(w)}}^{\dagger}\right]\delta^{\alpha}_{\beta}\delta^{a}_{b},
\end{equation}
and for magnetically coupled photons by
\begin{equation}
\frac{1}{2}\mathrm{Tr}\left[P_{j}^{(x)}\left(\kappa_{0}+\tau_{3}\kappa_{1}\right)\sigma_{n}{P_{i}^{(w)}}^{\dagger}\right]\delta^{\alpha}_{\beta}\delta^{a}_{b}.
\end{equation}
Projecting onto the doublet $S$-wave channel yields
\begin{align}
&\frac{1}{3}\left(\begin{array}{cc}
\sigma_{j}  & 0 \\[-1.5 mm]
0 & \tau_{B} 
\end{array}\right)
\left(\begin{array}{cc}
\frac{1}{2}\mathrm{Tr}\left[\sigma_{j}\sigma_{i}\left(\frac{1+\tau_{3}}{2}\right)\right] & \frac{1}{2}\mathrm{Tr}\left[\sigma_{j}\left(\frac{1+\tau_{3}}{2}\right)\tau_{A}\right] \\[0 mm] 
\frac{1}{2}\mathrm{Tr}\left[\sigma_{i}\tau_{B}\left(\frac{1+\tau_{3}}{2}\right)\right] & \frac{1}{2}\mathrm{Tr}\left[\tau_{B}\left(\frac{1+\tau_{3}}{2}\right)\tau_{A}\right]
\end{array}\right)
\left(\begin{array}{cc}
\sigma_{i}  & 0 \\[-1.5 mm]
0 & \tau_{A} 
\end{array}\right)=\\\nonumber
&\left(\begin{array}{cc}
1  & 0  \\[-1.5 mm]
0 & 1+\frac{2}{3}\tau_{3} 
\end{array}\right),
\end{align}
for minimally coupled $\hat{A}_{0}$ photons and
\begin{align}
&\frac{1}{3}\left(\begin{array}{cc}
\sigma_{j}  & 0 \\[-1.5 mm]
0 & \tau_{B} 
\end{array}\right)
\left(\begin{array}{cc}
\frac{1}{2}\mathrm{Tr}\left[\sigma_{j}\sigma_{n}\sigma_{i}(\kappa_{0}+\tau_{3}\kappa_{1})\right] & \frac{1}{2}\mathrm{Tr}\left[\sigma_{j}\sigma_{n}(\kappa_{0}+\tau_{3}\kappa_{1})\tau_{A}\right] \\[0 mm] 
\frac{1}{2}\mathrm{Tr}\left[\sigma_{n}\sigma_{i}\tau_{B}(\kappa_{0}+\tau_{3}\kappa_{1})\right] & \frac{1}{2}\mathrm{Tr}\left[\sigma_{n}\tau_{B}(\kappa_{0}+\tau_{3}\kappa_{1})\tau_{A}\right]
\end{array}\right)
\left(\begin{array}{cc}
\sigma_{i}  & 0 \\[-1.5 mm]
0 & \tau_{A} 
\end{array}\right)=\\\nonumber
&\left(\begin{array}{cc}
\frac{4}{3}\kappa_{0} & \frac{2}{3}\kappa_{1}\tau_{3} \\[-1.5 mm]
\frac{2}{3}\kappa_{1}\tau_{3} & 0
\end{array}\right)\sigma_{n},
\end{align}
for magnetically coupled photons.

The spin isospin c.c.~space operator for diagram-(d) of Fig.~\ref{fig:FormFactorNLO} for $\hat{A}_{0}$ photons from gauging the dibaryon kinetic term is given by
\begin{equation}
\delta_{wx}(c_{0t}^{(0)}\delta_{wt}\delta_{ij}+c_{0s}^{(0)}\delta_{ws}(2\delta_{i1}\delta_{j1}+\delta_{i0}\delta_{j0}))\delta^{\alpha}_{\beta}\delta^{a}_{b}.
\end{equation}
$\delta_{wt}$ picks out the contribution from the spin-triplet dibaryon and $\delta_{ws}$ from the spin-singlet dibaryon.  The indices $i$ and $j$ in $\delta_{i0}\delta_{j0}$ and $\delta_{i1}\delta_{j1}$  are spherical isospin indices and correspond to the fact that only the the $np$ and $pp$ spin-singlet dibaryon are charged and not the $nn$ spin-singlet dibaryon.  The spin isospin c.c.~space operator for diagram-(d) of Fig.~\ref{fig:FormFactorNLO} for the magnetic form factor comes the two-body currents in Eq.~(\ref{eq:2BLagMag}) which give
\begin{equation}
iL_{2}\epsilon^{jin}\delta_{wt}\delta_{xt}\delta^{\alpha}_{\beta}\delta^{a}_{b}-L_{1}(\delta_{ni}\delta_{j3}\delta_{wt}\delta_{xs}+\delta_{nj}\delta_{i3}\delta_{ws}\delta_{xt})\delta^{\alpha}_{\beta}\delta^{a}_{b}.
\end{equation}
Projecting in the doublet $S$-wave channel gives 
\begin{align}
\frac{1}{3}\left(\begin{array}{cc}
\sigma_{j}  & 0 \\[-1.5 mm]
0 & \tau_{B} 
\end{array}\right)
\left(\begin{array}{cc}
\delta_{ij}c_{0t}^{(0)} & 0 \\[0 mm] 
0 & (2\delta_{A1}+\delta_{A0})\delta_{AB}c_{0s}^{(0)}
\end{array}\right)
\left(\begin{array}{cc}
\sigma_{i}  & 0 \\[-1.5 mm]
0 & \tau_{A} 
\end{array}\right)=\left(\begin{array}{cc}
c_{0t}^{(0)}  & 0  \\[-1.5 mm]
0 &  (1+\frac{2}{3}\tau_{3})c_{0s}^{(0)}
\end{array}\right),
\end{align}
for $\hat{A}_{0}$ photons coupled to the dibaryons and
\begin{align}
\frac{1}{3}\left(\begin{array}{cc}
\sigma_{j}  & 0 \\[-1.5 mm]
0 & \tau_{B} 
\end{array}\right)
\left(\begin{array}{cc}
i\epsilon^{jin}L_{2} & -L_{1}\delta_{jn}\delta_{A3} \\[0 mm] 
-L_{1}\delta_{in}\delta_{B3} & 0
\end{array}\right)
\left(\begin{array}{cc}
\sigma_{i}  & 0 \\[-1.5 mm]
0 & \tau_{A} 
\end{array}\right)=\left(\begin{array}{cc}
-\frac{2}{3}L_{2} & -\frac{1}{3}L_{1}\tau_{3} \\[-1.5 mm]
-\frac{1}{3}L_{1}\tau_{3} & 0
\end{array}\right)\sigma_{n},
\end{align}
for photons magnetically coupled via the $L_{1}$ and $L_{2}$ two-body currents.
\section{\label{app:QExpansion}}
The scalar function $\mathcal{A}_{n}$ is given in Ref.~\cite{Vanasse:2015fph} for a specific choice of $a_{11}$ and $a_{22}$.  Generalizing to arbitrary $a_{11}$ and $a_{22}$ gives
\begin{align}
\mathcal{A}_{n}=\int_{0}^{\Lambda}dq q^{2}\left(a_{11}f^{(n)}_{t}(q)+a_{22}f^{(n)}_{s}(q)\right),
\end{align}
where
\begin{align}
f_{\{t,s\}}^{(0)}(q)=\frac{M_{N}}{384\pi^{2}}\frac{1}{\Dwd^{5}D_{\{t,s\}}^{4}}\left\{q^{2}(D_{\{t,s\}}^{2}-2D_{\{t,s\}}\Dwd+2\Dwd^{2})+4D_{\{t,s\}}\Dwd^{2}(3\Dwd-\gamma_{\{t,s\}})\right\},
\end{align}
\begin{equation}
f_{\{t,s\}}^{(1)}(q)=(Z_{\{t,s\}}-1)f_{\{t,s\}}^{(0)}(q),
\end{equation}
and
\begin{align}
&f_{\{t,s\}}^{(2)}(q)=\left(\frac{Z_{\{t,s\}}-1}{2\gamma_{\{t,s\}}}\right)^{2}\left[\vphantom{+\frac{M_{N}}{192\pi^{2}\Dwd^{3}D_{\{t,s\}}^{3}}\left\{8\Dwd^{2}D_{\{t,s\}}-q^{2}(\gamma_{\{t,s\}}-3\Dwd)\right\}}\left(\Dwd^{2}-\gamma_{\{t,s\}}^{2}\right)f_{\{t,s\}}^{(0)}(q)\right.\\\nonumber
&\hspace{5cm}\left.+\frac{M_{N}}{192\pi^{2}\Dwd^{3}D_{\{t,s\}}^{3}}\left\{8\Dwd^{2}D_{\{t,s\}}-q^{2}(\gamma_{\{t,s\}}-3\Dwd)\right\}\right].
\end{align}
The variables $D_{\{t,s\}}$ and $\Dwd$ are given by
\begin{equation}
\Dwd=\sqrt{\frac{3}{4}q^{2}-M_{N}E}\quad,\quad D_{\{t,s\}}=\gamma_{\{t,s\}}-\Dwd,
\end{equation}
where $\{t,s\}$ is a shorthand for two different functions one with subscript $t$ and the other with subscript $s$.  The c.c.~space vector function $\boldsymbol{\mathcal{A}}_{n}(p)$ is given by
\begin{equation}
\boldsymbol{\mathcal{A}}_{n}(p)=\int_{0}^{\Lambda}dq q^{2}
\left(\begin{array}{c}
a_{11}f_{t}^{(n)}(p,q)+3a_{22}f_{s}^{(n)}(p,q) \\
-a_{22}f_{s}^{(n)}(p,q)-3a_{11}f_{t}^{(n)}(p,q)
\end{array}\right),
\end{equation}
where
\begin{align}
&f_{\{t,s\}}^{(0)}(p,q)=-2\pi f_{\{t,s\}}^{(0)}(q)\frac{1}{p q}Q_{0}(a)\\\nonumber
&\hspace{1cm}-\frac{M_{N}}{27\pi}\frac{1}{D_{\{t,s\}}}\frac{1}{(pq)^{3}}\left\{\frac{5a}{(1-a^{2})^2}+\left[\left(\frac{q}{p}+\frac{p}{q}\right)(1+3a^{2})-a(3+a^{2})\right]\frac{1}{(1-a^{2})^{3}}\right\}\\\nonumber
&\hspace{1cm}-\frac{M_{N}}{432\pi}\frac{1}{\Dwd^{3}D_{\{t,s\}}^{3}}\frac{1}{(pq)^{2}}\left\{\Dwd^{2}D_{\{t,s\}}\left[\frac{38}{1-a^{2}}+\left(\left(20\frac{q}{p}+8\frac{p}{q}\right)a-4(1+a^{2})\right)\frac{1}{(1-a^{2})^{2}}\right]\right.\\\nonumber
&\hspace{1cm}\left.-(\gamma_{\{t,s\}}-3\Dwd)\frac{9}{2}\frac{q^{2}}{1-a^{2}}\right\},
\end{align}
\begin{align}
f_{\{t,s\}}^{(1)}(p,q)=&\left(\frac{Z_{\{t,s\}}-1}{2\gamma_{\{t,s\}}}\right)\left[(\gamma_{\{t,s\}}+\Dwd)f_{\{t,s\}}^{(0)}(p,q)-2\pi D_{\{t,s\}}f_{\{t,s\}}^{(0)}(q)\frac{1}{pq}Q_{0}(a)\right.\\\nonumber
&-\frac{M_{N}}{432\pi}\frac{1}{\Dwd^{3}D_{\{t,s\}}^{2}}\frac{1}{(pq)^{2}}\left\{\left[38\Dwd^{2}D_{\{t,s\}}-\frac{9}{2}q^{2}(\gamma_{\{t,s\}}-3\Dwd)\right]\frac{1}{1-a^{2}}\right.\\\nonumber
&\left.\left.-\Dwd^{2}D_{\{t,s\}}\left[4(1+a^{2})-\left(20\frac{q}{p}+8\frac{p}{q}\right)a\right]\frac{1}{(1-a^{2})^{2}}\right\}\right],
\end{align}
and
\begin{align}
f_{\{t,s\}}^{(2)}(p,q)=&\left(\frac{Z_{\{t,s\}}-1}{2\gamma_{\{t,s\}}}\right)^{2}\left[(\Dwd^{2}-\gamma_{\{t,s\}}^{2})f_{\{t,s\}}^{(0)}(p,q)\right.\\\nonumber
&-\frac{M_{N}}{96\pi}\frac{1}{\Dwd^{3}D_{\{t,s\}}^{3}}\left\{8\Dwd^{2}D_{\{t,s\}}-q^{2}(\gamma_{\{t,s\}}-3\Dwd)\right\}\frac{1}{pq}Q_{0}(a)\\\nonumber
&-\frac{M_{N}}{216\pi}\frac{1}{\Dwd D_{\{t,s\}}^{2}}\frac{1}{(pq)^{2}}\left\{\left[38\Dwd D_{\{t,s\}}+9q^{2}\right]\frac{1}{1-a^{2}}\right.\\\nonumber
&\left.\left.-\Dwd D_{\{t,s\}}\left[4(1+a^{2})-\left(20\frac{q}{p}+8\frac{p}{q}\right)a\right]\frac{1}{(1-a^{2})^{2}}\right\}\right].
\end{align}
The variable $a$ is defined by
\begin{equation}
a=\frac{q^{2}+p^{2}-M_{N}E}{qp}.
\end{equation}
$\boldsymbol{\mathcal{A}}_{n}(p,k)$ is c.c.~space matrix given by
\begin{align}
&\boldsymbol{\mathcal{A}}_{n}(p,k)=\\\nonumber
&\hspace{.3cm}\int_{0}^{\Lambda}dq q^{2}
\left(\begin{array}{cc}
a_{11}f_{t}^{(n)}(p,k,q)+9a_{22}f_{s}^{(n)}(p,k,q) & -3(a_{11}f_{t}^{(n)}(p,k,q)+a_{22}f_{s}^{(n)}(p,k,q)) \\
-3(a_{11}f_{t}^{(n)}(p,k,q)+a_{22}f_{s}^{(n)}(p,k,q)) & a_{22}f_{s}^{(n)}(p,k,q)+9a_{11}f_{t}^{(n)}(p,k,q)
\end{array}\right),
\end{align}
where
\begin{align}
f_{\{t,s\}}^{(0)}&(p,k,q)=-2\pi\left\{f_{\{t,s\}}^{(0)}(k,q)\frac{1}{pq}Q_{0}(a)+f_{\{t,s\}}^{(0)}(p,q)\frac{1}{kq}Q_{0}(b)\right\}\\\nonumber
&-4\pi^{2}f_{\{t,s\}}^{(0)}(q)\frac{1}{kq}Q_{0}(b)\frac{1}{pq}Q_{0}(a)\\\nonumber
&+\frac{M_{N}}{54}\frac{1}{\Dwd D_{\{t,s\}}^{2}}\frac{1}{q^{4}k^{2}p^{2}}\left\{2\Dwd D_{\{t,s\}} \left(\left[12(1-b^{2})(1-a^{2})+4\frac{q}{p}a(1-b^{2})+4\frac{q}{k}b(1-a^{2})\right]\right.\right.\\\nonumber
&+2ab\left[\frac{k}{p}(1-b^{2})+\frac{p}{k}(1-a^{2})\right]+2b\frac{k}{q}\left[2b^{2}-(1+a^{2})\right]+2a\frac{p}{q}\left[2a^{2}-(1+b^{2})\right]\\\nonumber
&\left.+2\frac{k}{q}\left(\frac{q}{p}a-2\right)(1-b^{2})^{2}Q_{0}(b)+2\frac{p}{q}\left(\frac{q}{k}b-2\right)(1-a^{2})^{2}Q_{0}(a)\right)\frac{1}{(1-b^{2})^{2}(1-a^{2})^{2}}\\\nonumber
&+q^{2}\left(\left[4+\frac{k}{q}b+\frac{p}{q}a-2\frac{k}{q}\frac{p}{q}ab\right]+\frac{k}{q}(1-b^{2})\left(1-2a\frac{p}{q}\right)Q_{0}(b)\right.\\\nonumber
&\left.\left.+\frac{p}{q}(1-a^{2})\left(1-2b\frac{k}{q}\right)Q_{0}(a)-2\frac{k}{q}\frac{p}{q}(1-b^{2})(1-a^{2})Q_{0}(b)Q_{0}(a)\right)\frac{1}{(1-b^{2})^{2}(1-a^{2})^{2}}\right\},
\end{align}
\begin{align}
f_{\{t,s\}}^{(1)}(p,k,q)=&\left(\frac{Z_{\{t,s\}}-1}{2\gamma_{\{t,s\}}}\right)(\gamma_{\{t,s\}}+\Dwd)f_{\{t,s\}}^{(0)}(p,k,q)\\\nonumber
&-2\pi f_{\{t,s\}}^{(1)}(k,q)\frac{1}{pq}Q_{0}(a)-2\pi f_{\{t,s\}}^{(1)}(p,q)\frac{1}{kq}Q_{0}(b)\\\nonumber
&+\left(\frac{Z_{\{t,s\}}-1}{2\gamma_{\{t,s\}}}\right)\frac{M_{N}}{54}\frac{1}{\Dwd D_{\{t,s\}}}\frac{1}{q^{2}k^{2}p^{2}}\left\{\left[4+\frac{k}{q}b+\frac{p}{q}a-2\frac{k}{q}\frac{p}{q}ab\right]\right.\\\nonumber
&+\frac{k}{q}(1-b^{2})\left(1-2a\frac{p}{q}\right)Q_{0}(b)+\frac{p}{q}(1-a^{2})\left(1-2b\frac{k}{q}\right)Q_{0}(a)\\\nonumber
&\left.\left.-2\frac{k}{q}\frac{p}{q}(1-b^{2})(1-a^{2})Q_{0}(b)Q_{0}(a)\right\}\frac{1}{(1-b^{2})(1-a^{2})}\right.\\\nonumber
&+2\pi \left(\frac{Z_{\{t,s\}}-1}{2\gamma_{\{t,s\}}}\right)(\gamma_{\{t,s\}}+\Dwd)\left[f_{\{t,s\}}^{(0)}(k,q)\frac{1}{pq}Q_{0}(a)+f_{\{t,s\}}^{(0)}(p,q)\frac{1}{kq}Q_{0}(b)\right]\\\nonumber
&-4\pi^{2}\left(f_{\{t,s\}}^{(1)}(q)-\left(\frac{Z_{\{t,s\}}-1}{2\gamma_{\{t,s\}}}\right)(\gamma_{\{t,s\}}+\Dwd)f_{\{t,s\}}^{(0)}(q)\right)\frac{1}{pq}Q_{0}(a)\frac{1}{kq}Q_{0}(b),
\end{align}
and
\begin{align}
f_{\{t,s\}}^{(2)}(p,k,q)=&\left(\frac{Z_{\{t,s\}}-1}{2\gamma_{\{t,s\}}}\right)^{2}(\Dwd^{2}-\gamma_{\{t,s\}}^{2})f_{\{t,s\}}^{(0)}(p,k,q)\\\nonumber
&-2\pi f_{\{t,s\}}^{(2)}(k,q)\frac{1}{pq}Q_{0}(a)-2\pi f_{\{t,s\}}^{(2)}(p,q)\frac{1}{kq}Q_{0}(b)\\\nonumber
&+\left(\frac{Z_{\{t,s\}}-1}{2\gamma_{\{t,s\}}}\right)^{2}\frac{M_{N}}{27}\frac{1}{D_{\{t,s\}}}\frac{1}{q^{2}k^{2}p^{2}}\left\{\left[4+\frac{k}{q}b+\frac{p}{q}a-2\frac{k}{q}\frac{p}{q}ab\right]\right.\\\nonumber
&+\frac{k}{q}(1-b^{2})\left(1-2a\frac{p}{q}\right)Q_{0}(b)+\frac{p}{q}(1-a^{2})\left(1-2b\frac{k}{q}\right)Q_{0}(a)\\\nonumber
&\left.-2\frac{k}{q}\frac{p}{q}(1-b^{2})(1-a^{2})Q_{0}(b)Q_{0}(a)\right\}\frac{1}{(1-b^{2})(1-a^{2})}\\\nonumber
&+2\pi\left(\frac{Z_{\{t,s\}}-1}{2\gamma_{\{t,s\}}}\right)^{2}(\Dwd^{2}-\gamma_{\{t,s\}}^{2})\left[f_{\{t,s\}}^{(0)}(k,q)\frac{1}{pq}Q_{0}(a)+f_{\{t,s\}}^{(0)}(p,q)\frac{1}{kq}Q_{0}(b)\right]\\\nonumber
&-4\pi^{2}\left(f_{\{t,s\}}^{(2)}(q)-\left(\frac{Z_{\{t,s\}}-1}{2\gamma_{\{t,s\}}}\right)^{2}(\Dwd^{2}-\gamma_{\{t,s\}}^{2})f_{\{t,s\}}^{(0)}(q)\right)\frac{1}{pq}Q_{0}(a)\frac{1}{kq}Q_{0}(b).
\end{align}
The variable $b$ is defined as
\begin{equation}
b=\frac{q^{2}+k^{2}-M_{N}E}{qk}.
\end{equation}

The c.c.~space matrix $\boldsymbol{\mathcal{B}}_{0}(p,k,Q)$ is given by
\begin{align}
&\boldsymbol{\mathcal{B}}_{0}(p,k)=-\frac{2M_{N}\pi}{9}\frac{1}{p^{3}k^{3}}\frac{1}{(1-a^{2})^{2}}\\\nonumber
&\hspace{4cm}\times\left\{\frac{4}{3}\frac{a}{1-a^{2}}-2a-\frac{1}{3}\frac{p^{2}+k^{2}}{pk}\frac{1+3a^{2}}{1-a^{2}}\right\}\left(\!\!\begin{array}{rr}
b_{11} & b_{12}\\
b_{21} & b_{22}
\end{array}\!\right),
\end{align}
where now and for the rest of the appendix
\begin{equation}
a=\frac{p^{2}+k^{2}-M_{N}E}{pk}.
\end{equation}

$\boldsymbol{\mathcal{C}}_{n}(k)$ is a c.c.~space vector given by
\begin{align}
\boldsymbol{\mathcal{C}}_{n}(k)=
\left(\begin{array}{c}
2c_{11}g_{t}^{(n)}(k)-2c_{21}g_{s}^{(n)}(k) \\
2c_{12}g_{t}^{(n)}(k)-2c_{22}g_{s}^{(n)}(k)
\end{array}\right)^{T},
\end{align}
where
\begin{align}
g_{\{t,s\}}^{(0)}(k)=\frac{M_{N}}{384\Dwd^{5}D_{\{t,s\}}^{3}}\left\{4\Dwd^{2}D_{\{t,s\}}(2\Dwd-\gamma_{\{t,s\}})+k^{2}(\gamma_{\{t,s\}}-3\Dwd)D_{\{t,s\}}+2k^{2}\Dwd^{2}\right\},
\end{align}
\begin{align}
g_{\{t,s\}}^{(1)}(k)=&\left(\frac{Z_{\{t,s\}}-1}{2\gamma_{\{t,s\}}}\right)\left[\vphantom{+\frac{M_{N}}{192\Dwd^{4}D_{\{t,s\}}^{2}}\left\{2\Dwd^{2} D_{\{t,s\}}+k^{2}(\Dwd-D_{\{t,s\}})\right\}}(\gamma_{\{t,s\}}+\Dwd)g_{\{t,s\}}^{(0)}(k)\right.\\\nonumber
&\hspace{3cm}\left.+\frac{M_{N}}{192\Dwd^{4}D_{\{t,s\}}^{2}}\left\{2\Dwd^{2} D_{\{t,s\}}+k^{2}(\Dwd-D_{\{t,s\}})\right\}\right],
\end{align}
and 
\begin{align}
g_{\{t,s\}}^{(2)}(k)=&\left(\frac{Z_{\{t,s\}}-1}{2\gamma_{\{t,s\}}}\right)^{2}\left[\vphantom{+\frac{M_{N}}{96\Dwd^{3}D_{\{t,s\}}^{2}}\left\{2\Dwd^{2}D_{\{t,s\}}+k^{2}\left(\Dwd-\frac{1}{2}D_{\{t,s\}}\right)\right\}}(\Dwd^{2}-\gamma_{\{t,s\}}^{2})g_{\{t,s\}}^{(0)}(k)\right.\\\nonumber
&\hspace{3cm}\left.+\frac{M_{N}}{96\Dwd^{3}D_{\{t,s\}}^{2}}\left\{2\Dwd^{2}D_{\{t,s\}}+k^{2}\left(\Dwd-\frac{1}{2}D_{\{t,s\}}\right)\right\}\right].
\end{align}
For these functions and all functions below in this appendix the variables $\Dwd$, $D_{t}$, and $D_{s}$ are defined as
\begin{equation}
\Dwd=\sqrt{\frac{3}{4}k^{2}-M_{N}E}\quad,\quad D_{\{t,s\}}=\gamma_{\{t,s\}}-\Dwd
\end{equation}

The c.c.~space matrix $\boldsymbol{\mathcal{C}}_{n}(p,k)$ is defined by
\begin{align}
&\boldsymbol{\mathcal{C}}_{n}(p,k)=\left(
\begin{array}{cc}
2\left[c_{11}g_{t}^{(n)}(p,k)-3c_{21}g_{s}^{(n)}(p,k)\right] & 2\left[c_{12}g_{t}^{(n)}(p,k)-3c_{22}g_{s}^{(n)}(p,k)\right]\\[2mm]
2\left[c_{21}g_{s}^{(n)}(p,k)-3c_{11}g_{t}^{(n)}(p,k)\right] & 2\left[c_{22}g_{s}^{(n)}(p,k)-3c_{12}g_{t}^{(n)}(p,k)\right]
\end{array}\right),
\end{align}
where
\begin{align}
g_{\{t,s\}}^{(0)}(p,k)=&-2\pi g_{\{t,s\}}^{(0)}(k)\frac{1}{pk}Q_{0}(a)\\\nonumber
&-\frac{M_{N}\pi}{54\Dwd D_{\{t,s\}}}\frac{1}{pk}\left\{\frac{1}{pk}\frac{1}{1-a^{2}}+\frac{1}{p^{2}}\left(4a+a\left(\frac{p}{k}\right)^{2}-2\frac{p}{k}(1+a^{2})\right)\frac{1}{(1-a^{2})^{2}}\right\}\\\nonumber
&-\frac{M_{N}\pi}{144}\frac{k}{p}\frac{1}{\Dwd^{3}D_{\{t,s\}}^{2}}\left\{\frac{1}{k^{2}}Q_{0}(a)-\frac{1}{pk}\frac{2-\frac{p}{k}a}{1-a^{2}}\right\}\left[\gamma_{\{t,s\}}-3\Dwd\right],
\end{align}
\begin{align}
&g_{\{t,s\}}^{(1)}(p,k)=\left(\frac{Z_{\{t,s\}}-1}{2\gamma_{t}}\right)\left[\vphantom{\left\{\frac{2}{pk}\frac{1}{1-a^{2}}-\frac{1}{k^{2}}\frac{a}{1-a^{2}}-\frac{1}{k^{2}}Q_{0}(a)\right\}}(\gamma_{\{t,s\}}+\Dwd)g_{\{t,s\}}^{(0)}(p,k)\right.\\\nonumber
&\hspace{2cm}-\frac{M_{N}\pi}{96\Dwd^{4} D_{\{t,s\}}^{2}}\frac{1}{pk}Q_{0}(a)\left\{2\Dwd^{2} D_{\{t,s\}}+k^{2}(\Dwd-D_{\{t,s\}})\right\}\\\nonumber
&\hspace{2cm}\left.-\frac{k}{p}\frac{M_{N}\pi}{72\Dwd^{2}D_{\{t,s\}}}\left\{\frac{2}{pk}\frac{1}{1-a^{2}}-\frac{1}{k^{2}}\frac{a}{1-a^{2}}-\frac{1}{k^{2}}Q_{0}(a)\right\}\right],
\end{align}
and
\begin{align}
g_{\{t,s\}}^{(2)}(p,k)=&\left(\frac{Z_{t}-1}{2\gamma_{t}}\right)^{2}\left[(\Dwd^{2}-\gamma_{\{t,s\}}^{2})g_{\{t,s\}}^{(0)}(p,k)\right.\\\nonumber
&-\frac{M_{N}\pi}{48\Dwd^{3} D_{\{t,s\}}^{2}}\frac{1}{pk}Q_{0}(a)\left\{2\Dwd^{2}D_{\{t,s\}}+k^{2}\left(\Dwd-\frac{1}{2}D_{\{t,s\}}\right)\right\}\\\nonumber
&\left.-\frac{k}{p}\frac{M_{N}\pi}{36\Dwd D_{\{t,s\}}}\left\{\frac{2}{pk}\frac{1}{1-a^{2}}-\frac{1}{k^{2}}\frac{a}{1-a^{2}}-\frac{1}{k^{2}}Q_{0}(a)\right\}\right].
\end{align}

$\boldsymbol{\mathfrak{D}}_{n}(k)$ is a c.c~space vector given by
\begin{align}
\boldsymbol{\mathfrak{D}}_{n}(k)=\left(\begin{array}{c}d_{11}h_{t}^{(n)}(k)-d_{21}h_{s}^{(n)}(k) \\[2mm]
d_{12}h_{t}^{(n)}(k)-d_{22}h_{s}^{(n)}(k)
\end{array}\right)^{T},
\end{align}
where
\begin{align}
h_{\{t,s\}}^{(1)}(k)=-\frac{1}{96\Dwd^{3}D_{\{t,s\}}^{3}}\left\{4\Dwd^{2}D_{\{t,s\}}+k^{2}(3\Dwd-\gamma_{\{t,s\}})\right\},
\end{align}
and
%
%
\begin{align}
h_{\{t,s\}}^{(2)}(k)=0.
\end{align}
Finally, the c.c.~space matrix $\boldsymbol{\mathfrak{D}}_{n}(p,k)$ is given by
\begin{align}
&\boldsymbol{\mathfrak{D}}_{n}(p,k)=
\left(\begin{array}{cc}
\left[d_{11}h_{t}^{(n)}(p,k)-3d_{21}h_{s}^{(n)}(p,k)\right] & \left[d_{12}^{(0)}h_{t}^{(n)}(p,k)-3d_{22}h_{s}^{(n)}(p,k)\right]\\[2mm]
\left[d_{21}h_{s}^{(n)}(p,k)-3d_{11}h_{t}^{(n)}(p,k)\right] & \left[d_{22}h_{s}^{(n)}(p,k)-3d_{12}^{(0)}h_{t}^{(n)}(p,k)\right]\\
\end{array}\right),
\end{align}
where
\begin{align}
h_{\{t,s\}}^{(1)}(p,k)=&-2\pi h_{\{t,s\}}^{(1)}(k)\frac{1}{pk}Q_{0}(a)\\\nonumber
&+\frac{2\pi}{27 D_{\{t,s\}}}\frac{1}{(pk)^{2}}\left[\left(4\frac{k}{p}+\frac{p}{k}\right)a-3a^{2}-1\right]\frac{1}{(1-a^{2})^{2}}\\\nonumber
&-\frac{\pi}{18\Dwd D_{\{t,s\}}^{2}}\frac{1}{pk}\left\{Q_{0}(a)+\frac{a-2\frac{k}{p}}{1-a^{2}}\right\},
\end{align}
and
%
%
\begin{align}
&h_{\{t,s\}}^{(2)}(p,k)=-\left(\frac{Z_{\{t,s\}}-1}{2\gamma_{\{t,s\}}}\right)\left[D_{\{t,s\}}h_{\{t,s\}}^{(1)}(p,k)+2\pi D_{\{t,s\}}h_{\{t,s\}}^{(1)}(k)\frac{1}{pk}Q_{0}(a)\right.\\\nonumber
&\hspace{3cm}\left.-\frac{\pi}{18\Dwd D_{\{t,s\}}}\frac{1}{pk}\left\{\left[2\frac{k}{p}-a\right]\frac{1}{1-a^{2}}-Q_{0}(a)\right\}\right].
\end{align}
The $\nnlo$ functions $h_{\{t,s\}}^{(2)}(k)$ and $h_{\{t,s\}}^{(2)}(p,k)$ can only be used for the charge form factor.  These functions include $\nnlo$ corrections from gauging $c_{0t}^{(1)}$ and $c_{0s}^{(1)}$ that should not be included in the magnetic form factor.


\section{\label{app:Wigner} Wigner-SU(4) limit}

The three-nucleon magnetic moments are defined by the matrix element
\begin{equation}
\frac{e}{2M_{N}}\left<{}^{A}Z\Bigg{|}\sum_{i=1}^{3}(\kappa_{0}+\kappa_{1}\tau_{3}^{(i)})\sigma_{3}^{(i)}\Bigg{|}{}^{A}Z\right>,
\end{equation}
where the nucleon magnetic moment term is inserted between the three-nucleon wavefunction $\left|^{A}Z\right>$ and summed over all nucleons in the system.  Summing over all nucleons yields the operators
\begin{equation}
\sum_{i=1}^{3}\sigma_{3}^{(i)}=2S_{z}\quad,\quad \sum_{i=1}^{3}\tau_{3}^{i}\sigma_{3}^{(i)}=2Y_{zz},
\end{equation}
where $S_{z}$ is the total spin of the three-nucleon system and $Y_{zz}$ an SU(4) operator.  Assuming the three-nucleon system has Wigner-SU(4) symmetry than it is an eigenstate of the operator $Y_{zz}$ and the matrix element reduces to

\begin{equation}
\frac{e}{2M_{N}}\left<{}^{A}Z\Bigg{|}2(\kappa_{0}S_{z}+\kappa_{1}Y_{zz})\Bigg{|}{}^{A}Z\right>=
\left\{\begin{array}{ll}
\frac{e}{2M_{N}}(\kappa_{0}+\kappa_{1}), & {}^{A}Z=\jjvH\\
\frac{e}{2M_{N}}(\kappa_{0}-\kappa_{1}), & {}^{A}Z=\jjvHe
\end{array}\right..
\end{equation}


\begin{thebibliography}{68}
\expandafter\ifx\csname natexlab\endcsname\relax\def\natexlab#1{#1}\fi
\expandafter\ifx\csname bibnamefont\endcsname\relax
  \def\bibnamefont#1{#1}\fi
\expandafter\ifx\csname bibfnamefont\endcsname\relax
  \def\bibfnamefont#1{#1}\fi
\expandafter\ifx\csname citenamefont\endcsname\relax
  \def\citenamefont#1{#1}\fi
\expandafter\ifx\csname url\endcsname\relax
  \def\url#1{\texttt{#1}}\fi
\expandafter\ifx\csname urlprefix\endcsname\relax\def\urlprefix{URL }\fi
\providecommand{\bibinfo}[2]{#2}
\providecommand{\eprint}[2][]{\url{#2}}

\bibitem[{\citenamefont{Angeli and Marinova}(2013)}]{Angeli201369}
\bibinfo{author}{\bibfnamefont{I.}~\bibnamefont{Angeli}} \bibnamefont{and}
  \bibinfo{author}{\bibfnamefont{K.}~\bibnamefont{Marinova}},
  \bibinfo{journal}{Atomic Data and Nuclear Data Tables}
  \textbf{\bibinfo{volume}{99}}, \bibinfo{pages}{69 } (\bibinfo{year}{2013}),
  ISSN \bibinfo{issn}{0092-640X},
  \urlprefix\url{http://www.sciencedirect.com/science/article/pii/S0092640X12000265}.

\bibitem[{\citenamefont{Sick}(2001)}]{Sick:2001rh}
\bibinfo{author}{\bibfnamefont{I.}~\bibnamefont{Sick}}, \bibinfo{journal}{Prog.
  Part. Nucl. Phys.} \textbf{\bibinfo{volume}{47}}, \bibinfo{pages}{245}
  (\bibinfo{year}{2001}), \eprint{nucl-ex/0208009}.

\bibitem[{\citenamefont{Braaten and Hammer}(2006)}]{Braaten:2004rn}
\bibinfo{author}{\bibfnamefont{E.}~\bibnamefont{Braaten}} \bibnamefont{and}
  \bibinfo{author}{\bibfnamefont{H.-W.} \bibnamefont{Hammer}},
  \bibinfo{journal}{Phys. Rept.} \textbf{\bibinfo{volume}{428}},
  \bibinfo{pages}{259} (\bibinfo{year}{2006}), \eprint{cond-mat/0410417}.

\bibitem[{\citenamefont{van Kolck}(1999)}]{vanKolck:1998bw}
\bibinfo{author}{\bibfnamefont{U.}~\bibnamefont{van Kolck}},
  \bibinfo{journal}{Nucl.Phys.} \textbf{\bibinfo{volume}{A645}},
  \bibinfo{pages}{273} (\bibinfo{year}{1999}), \eprint{nucl-th/9808007}.

\bibitem[{\citenamefont{Kaplan et~al.}(1998{\natexlab{a}})\citenamefont{Kaplan,
  Savage, and Wise}}]{Kaplan:1998tg}
\bibinfo{author}{\bibfnamefont{D.~B.} \bibnamefont{Kaplan}},
  \bibinfo{author}{\bibfnamefont{M.~J.} \bibnamefont{Savage}},
  \bibnamefont{and} \bibinfo{author}{\bibfnamefont{M.~B.} \bibnamefont{Wise}},
  \bibinfo{journal}{Phys. Lett. B} \textbf{\bibinfo{volume}{424}},
  \bibinfo{pages}{390} (\bibinfo{year}{1998}{\natexlab{a}}),
  \eprint{nucl-th/9801034}.

\bibitem[{\citenamefont{Kaplan et~al.}(1998{\natexlab{b}})\citenamefont{Kaplan,
  Savage, and Wise}}]{Kaplan:1998we}
\bibinfo{author}{\bibfnamefont{D.~B.} \bibnamefont{Kaplan}},
  \bibinfo{author}{\bibfnamefont{M.~J.} \bibnamefont{Savage}},
  \bibnamefont{and} \bibinfo{author}{\bibfnamefont{M.~B.} \bibnamefont{Wise}},
  \bibinfo{journal}{Nucl. Phys. B} \textbf{\bibinfo{volume}{534}},
  \bibinfo{pages}{329} (\bibinfo{year}{1998}{\natexlab{b}}),
  \eprint{nucl-th/9802075}.

\bibitem[{\citenamefont{Chen et~al.}(1999)\citenamefont{Chen, Rupak, and
  Savage}}]{Chen:1999tn}
\bibinfo{author}{\bibfnamefont{J.-W.} \bibnamefont{Chen}},
  \bibinfo{author}{\bibfnamefont{G.}~\bibnamefont{Rupak}}, \bibnamefont{and}
  \bibinfo{author}{\bibfnamefont{M.~J.} \bibnamefont{Savage}},
  \bibinfo{journal}{Nucl. Phys. A} \textbf{\bibinfo{volume}{653}},
  \bibinfo{pages}{386} (\bibinfo{year}{1999}), \eprint{nucl-th/9902056}.

\bibitem[{\citenamefont{Kong and Ravndal}(1999)}]{Kong:1998sx}
\bibinfo{author}{\bibfnamefont{X.}~\bibnamefont{Kong}} \bibnamefont{and}
  \bibinfo{author}{\bibfnamefont{F.}~\bibnamefont{Ravndal}},
  \bibinfo{journal}{Phys. Lett. B} \textbf{\bibinfo{volume}{450}},
  \bibinfo{pages}{320} (\bibinfo{year}{1999}), \eprint{nucl-th/9811076}.

\bibitem[{\citenamefont{Kong and Ravndal}(2000)}]{Kong:1999sf}
\bibinfo{author}{\bibfnamefont{X.}~\bibnamefont{Kong}} \bibnamefont{and}
  \bibinfo{author}{\bibfnamefont{F.}~\bibnamefont{Ravndal}},
  \bibinfo{journal}{Nucl. Phys. A} \textbf{\bibinfo{volume}{665}},
  \bibinfo{pages}{137} (\bibinfo{year}{2000}), \eprint{hep-ph/9903523}.

\bibitem[{\citenamefont{Ando et~al.}(2007)\citenamefont{Ando, Shin, Hyun, and
  Hong}}]{Ando:2007fh}
\bibinfo{author}{\bibfnamefont{S.-I.} \bibnamefont{Ando}},
  \bibinfo{author}{\bibfnamefont{J.~W.} \bibnamefont{Shin}},
  \bibinfo{author}{\bibfnamefont{C.~H.} \bibnamefont{Hyun}}, \bibnamefont{and}
  \bibinfo{author}{\bibfnamefont{S.~W.} \bibnamefont{Hong}},
  \bibinfo{journal}{Phys. Rev. C} \textbf{\bibinfo{volume}{76}},
  \bibinfo{pages}{064001} (\bibinfo{year}{2007}), \eprint{0704.2312}.

\bibitem[{\citenamefont{Chen and Savage}(1999)}]{Chen:1999bg}
\bibinfo{author}{\bibfnamefont{J.-W.} \bibnamefont{Chen}} \bibnamefont{and}
  \bibinfo{author}{\bibfnamefont{M.~J.} \bibnamefont{Savage}},
  \bibinfo{journal}{Phys. Rev. C} \textbf{\bibinfo{volume}{60}},
  \bibinfo{pages}{065205} (\bibinfo{year}{1999}), \eprint{nucl-th/9907042}.

\bibitem[{\citenamefont{Ando and Hyun}(2005)}]{Ando:2004mm}
\bibinfo{author}{\bibfnamefont{S.-I.} \bibnamefont{Ando}} \bibnamefont{and}
  \bibinfo{author}{\bibfnamefont{C.~H.} \bibnamefont{Hyun}},
  \bibinfo{journal}{Phys. Rev. C} \textbf{\bibinfo{volume}{72}},
  \bibinfo{pages}{014008} (\bibinfo{year}{2005}), \eprint{nucl-th/0407103}.

\bibitem[{\citenamefont{Rupak}(2000)}]{Rupak:1999rk}
\bibinfo{author}{\bibfnamefont{G.}~\bibnamefont{Rupak}},
  \bibinfo{journal}{Nucl. Phys. A} \textbf{\bibinfo{volume}{678}},
  \bibinfo{pages}{405} (\bibinfo{year}{2000}), \eprint{nucl-th/9911018}.

\bibitem[{\citenamefont{Kong and Ravndal}(2001)}]{Kong:2000px}
\bibinfo{author}{\bibfnamefont{X.}~\bibnamefont{Kong}} \bibnamefont{and}
  \bibinfo{author}{\bibfnamefont{F.}~\bibnamefont{Ravndal}},
  \bibinfo{journal}{Phys. Rev. C} \textbf{\bibinfo{volume}{64}},
  \bibinfo{pages}{044002} (\bibinfo{year}{2001}), \eprint{nucl-th/0004038}.

\bibitem[{\citenamefont{Ando et~al.}(2008)\citenamefont{Ando, Shin, Hyun, Hong,
  and Kubodera}}]{Ando:2008va}
\bibinfo{author}{\bibfnamefont{S.-I.} \bibnamefont{Ando}},
  \bibinfo{author}{\bibfnamefont{J.}~\bibnamefont{Shin}},
  \bibinfo{author}{\bibfnamefont{C.}~\bibnamefont{Hyun}},
  \bibinfo{author}{\bibfnamefont{S.}~\bibnamefont{Hong}}, \bibnamefont{and}
  \bibinfo{author}{\bibfnamefont{K.}~\bibnamefont{Kubodera}},
  \bibinfo{journal}{Phys. Lett. B} \textbf{\bibinfo{volume}{668}},
  \bibinfo{pages}{187} (\bibinfo{year}{2008}), \eprint{0801.4330}.

\bibitem[{\citenamefont{Chen et~al.}(2013)\citenamefont{Chen, Liu, and
  Yu}}]{Chen:2012hm}
\bibinfo{author}{\bibfnamefont{J.-W.} \bibnamefont{Chen}},
  \bibinfo{author}{\bibfnamefont{C.-P.} \bibnamefont{Liu}}, \bibnamefont{and}
  \bibinfo{author}{\bibfnamefont{S.-H.} \bibnamefont{Yu}},
  \bibinfo{journal}{Phys. Lett. B} \textbf{\bibinfo{volume}{720}},
  \bibinfo{pages}{385} (\bibinfo{year}{2013}), \eprint{1209.2552}.

\bibitem[{\citenamefont{Butler et~al.}(2001)\citenamefont{Butler, Chen, and
  Kong}}]{Butler:2000zp}
\bibinfo{author}{\bibfnamefont{M.}~\bibnamefont{Butler}},
  \bibinfo{author}{\bibfnamefont{J.-W.} \bibnamefont{Chen}}, \bibnamefont{and}
  \bibinfo{author}{\bibfnamefont{X.}~\bibnamefont{Kong}},
  \bibinfo{journal}{Phys. Rev. C} \textbf{\bibinfo{volume}{63}},
  \bibinfo{pages}{035501} (\bibinfo{year}{2001}), \eprint{nucl-th/0008032}.

\bibitem[{\citenamefont{Bedaque et~al.}(1998)\citenamefont{Bedaque, Hammer, and
  van Kolck}}]{Bedaque:1998mb}
\bibinfo{author}{\bibfnamefont{P.~F.} \bibnamefont{Bedaque}},
  \bibinfo{author}{\bibfnamefont{H.-W.} \bibnamefont{Hammer}},
  \bibnamefont{and} \bibinfo{author}{\bibfnamefont{U.}~\bibnamefont{van
  Kolck}}, \bibinfo{journal}{Phys. Rev. C} \textbf{\bibinfo{volume}{58}},
  \bibinfo{pages}{641} (\bibinfo{year}{1998}), \eprint{nucl-th/9802057}.

\bibitem[{\citenamefont{Bedaque et~al.}(2000)\citenamefont{Bedaque, Hammer, and
  van Kolck}}]{Bedaque:1999ve}
\bibinfo{author}{\bibfnamefont{P.~F.} \bibnamefont{Bedaque}},
  \bibinfo{author}{\bibfnamefont{H.-W.} \bibnamefont{Hammer}},
  \bibnamefont{and} \bibinfo{author}{\bibfnamefont{U.}~\bibnamefont{van
  Kolck}}, \bibinfo{journal}{Nucl. Phys. A} \textbf{\bibinfo{volume}{676}},
  \bibinfo{pages}{357} (\bibinfo{year}{2000}), \eprint{nucl-th/9906032}.

\bibitem[{\citenamefont{Gabbiani et~al.}(2000)\citenamefont{Gabbiani, Bedaque,
  and Grie{\ss}hammer}}]{Gabbiani:1999yv}
\bibinfo{author}{\bibfnamefont{F.}~\bibnamefont{Gabbiani}},
  \bibinfo{author}{\bibfnamefont{P.~F.} \bibnamefont{Bedaque}},
  \bibnamefont{and} \bibinfo{author}{\bibfnamefont{H.~W.}
  \bibnamefont{Grie{\ss}hammer}}, \bibinfo{journal}{Nucl. Phys. A}
  \textbf{\bibinfo{volume}{675}}, \bibinfo{pages}{601} (\bibinfo{year}{2000}),
  \eprint{nucl-th/9911034}.

\bibitem[{\citenamefont{Bedaque et~al.}(2003)\citenamefont{Bedaque, Rupak,
  Grie{\ss}hammer, and Hammer}}]{Bedaque:2002yg}
\bibinfo{author}{\bibfnamefont{P.~F.} \bibnamefont{Bedaque}},
  \bibinfo{author}{\bibfnamefont{G.}~\bibnamefont{Rupak}},
  \bibinfo{author}{\bibfnamefont{H.~W.} \bibnamefont{Grie{\ss}hammer}},
  \bibnamefont{and} \bibinfo{author}{\bibfnamefont{H.-W.}
  \bibnamefont{Hammer}}, \bibinfo{journal}{Nucl. Phys. A}
  \textbf{\bibinfo{volume}{714}}, \bibinfo{pages}{589} (\bibinfo{year}{2003}),
  \eprint{nucl-th/0207034}.

\bibitem[{\citenamefont{Grie{\ss}hammer}(2004)}]{Griesshammer:2004pe}
\bibinfo{author}{\bibfnamefont{H.~W.} \bibnamefont{Grie{\ss}hammer}},
  \bibinfo{journal}{Nucl. Phys. A} \textbf{\bibinfo{volume}{744}},
  \bibinfo{pages}{192} (\bibinfo{year}{2004}), \eprint{nucl-th/0404073}.

\bibitem[{\citenamefont{Vanasse}(2013)}]{Vanasse:2013sda}
\bibinfo{author}{\bibfnamefont{J.}~\bibnamefont{Vanasse}},
  \bibinfo{journal}{Phys. Rev. C} \textbf{\bibinfo{volume}{88}},
  \bibinfo{pages}{044001} (\bibinfo{year}{2013}), \eprint{1305.0283}.

\bibitem[{\citenamefont{Margaryan et~al.}(2016)\citenamefont{Margaryan,
  Springer, and Vanasse}}]{Margaryan:2015rzg}
\bibinfo{author}{\bibfnamefont{A.}~\bibnamefont{Margaryan}},
  \bibinfo{author}{\bibfnamefont{R.~P.} \bibnamefont{Springer}},
  \bibnamefont{and} \bibinfo{author}{\bibfnamefont{J.}~\bibnamefont{Vanasse}},
  \bibinfo{journal}{Phys. Rev.} \textbf{\bibinfo{volume}{C93}},
  \bibinfo{pages}{054001} (\bibinfo{year}{2016}), \eprint{1512.03774}.

\bibitem[{\citenamefont{Rupak and Kong}(2003)}]{Rupak:2001ci}
\bibinfo{author}{\bibfnamefont{G.}~\bibnamefont{Rupak}} \bibnamefont{and}
  \bibinfo{author}{\bibfnamefont{X.-w.} \bibnamefont{Kong}},
  \bibinfo{journal}{Nucl. Phys. A} \textbf{\bibinfo{volume}{717}},
  \bibinfo{pages}{73} (\bibinfo{year}{2003}), \eprint{nucl-th/0108059}.

\bibitem[{\citenamefont{K{\"o}nig and Hammer}(2011)}]{Konig:2011yq}
\bibinfo{author}{\bibfnamefont{S.}~\bibnamefont{K{\"o}nig}} \bibnamefont{and}
  \bibinfo{author}{\bibfnamefont{H.-W.} \bibnamefont{Hammer}},
  \bibinfo{journal}{Phys. Rev. C} \textbf{\bibinfo{volume}{83}},
  \bibinfo{pages}{064001} (\bibinfo{year}{2011}), \eprint{1101.5939}.

\bibitem[{\citenamefont{K{\"o}nig and Hammer}(2014)}]{Konig:2013cia}
\bibinfo{author}{\bibfnamefont{S.}~\bibnamefont{K{\"o}nig}} \bibnamefont{and}
  \bibinfo{author}{\bibfnamefont{H.-W.} \bibnamefont{Hammer}},
  \bibinfo{journal}{Phys. Rev.} \textbf{\bibinfo{volume}{C90}},
  \bibinfo{pages}{034005} (\bibinfo{year}{2014}), \eprint{1312.2573}.

\bibitem[{\citenamefont{Vanasse et~al.}(2014)\citenamefont{Vanasse, Egolf,
  Kerin, König, and Springer}}]{Vanasse:2014kxa}
\bibinfo{author}{\bibfnamefont{J.}~\bibnamefont{Vanasse}},
  \bibinfo{author}{\bibfnamefont{D.~A.} \bibnamefont{Egolf}},
  \bibinfo{author}{\bibfnamefont{J.}~\bibnamefont{Kerin}},
  \bibinfo{author}{\bibfnamefont{S.}~\bibnamefont{König}}, \bibnamefont{and}
  \bibinfo{author}{\bibfnamefont{R.~P.} \bibnamefont{Springer}},
  \bibinfo{journal}{Phys. Rev.} \textbf{\bibinfo{volume}{C89}},
  \bibinfo{pages}{064003} (\bibinfo{year}{2014}), \eprint{1402.5441}.

\bibitem[{\citenamefont{K{\" o}nig et~al.}(2015)\citenamefont{K{\" o}nig,
  Grießhammer, and Hammer}}]{Konig:2014ufa}
\bibinfo{author}{\bibfnamefont{S.}~\bibnamefont{K{\" o}nig}},
  \bibinfo{author}{\bibfnamefont{H.~W.} \bibnamefont{Grießhammer}},
  \bibnamefont{and} \bibinfo{author}{\bibfnamefont{H.-W.}
  \bibnamefont{Hammer}}, \bibinfo{journal}{J. Phys.}
  \textbf{\bibinfo{volume}{G42}}, \bibinfo{pages}{045101}
  (\bibinfo{year}{2015}), \eprint{1405.7961}.

\bibitem[{\citenamefont{K{\"o}nig}(2016)}]{Konig:2016iny}
\bibinfo{author}{\bibfnamefont{S.}~\bibnamefont{K{\"o}nig}}
  (\bibinfo{year}{2016}), \eprint{1609.03163}.

\bibitem[{\citenamefont{Ando and Birse}(2010)}]{Ando:2010wq}
\bibinfo{author}{\bibfnamefont{S.-I.} \bibnamefont{Ando}} \bibnamefont{and}
  \bibinfo{author}{\bibfnamefont{M.~C.} \bibnamefont{Birse}},
  \bibinfo{journal}{J. Phys. G} \textbf{\bibinfo{volume}{37}},
  \bibinfo{pages}{105108} (\bibinfo{year}{2010}), \eprint{1003.4383}.

\bibitem[{\citenamefont{König et~al.}(2016)\citenamefont{König, Grießhammer,
  Hammer, and van Kolck}}]{Konig:2015aka}
\bibinfo{author}{\bibfnamefont{S.}~\bibnamefont{König}},
  \bibinfo{author}{\bibfnamefont{H.~W.} \bibnamefont{Grießhammer}},
  \bibinfo{author}{\bibfnamefont{H.-W.} \bibnamefont{Hammer}},
  \bibnamefont{and} \bibinfo{author}{\bibfnamefont{U.}~\bibnamefont{van
  Kolck}}, \bibinfo{journal}{J. Phys.} \textbf{\bibinfo{volume}{G43}},
  \bibinfo{pages}{055106} (\bibinfo{year}{2016}), \eprint{1508.05085}.

\bibitem[{\citenamefont{Platter and Hammer}(2006)}]{Platter:2005sj}
\bibinfo{author}{\bibfnamefont{L.}~\bibnamefont{Platter}} \bibnamefont{and}
  \bibinfo{author}{\bibfnamefont{H.-W.} \bibnamefont{Hammer}},
  \bibinfo{journal}{Nucl. Phys.} \textbf{\bibinfo{volume}{A766}},
  \bibinfo{pages}{132} (\bibinfo{year}{2006}), \eprint{nucl-th/0509045}.

\bibitem[{\citenamefont{Kirscher et~al.}(2017)\citenamefont{Kirscher, Pazy,
  Drachman, and Barnea}}]{Kirscher:2017fqc}
\bibinfo{author}{\bibfnamefont{J.}~\bibnamefont{Kirscher}},
  \bibinfo{author}{\bibfnamefont{E.}~\bibnamefont{Pazy}},
  \bibinfo{author}{\bibfnamefont{J.}~\bibnamefont{Drachman}}, \bibnamefont{and}
  \bibinfo{author}{\bibfnamefont{N.}~\bibnamefont{Barnea}}
  (\bibinfo{year}{2017}), \eprint{1702.07268}.

\bibitem[{\citenamefont{De-Leon et~al.}(2016)\citenamefont{De-Leon, Platter,
  and Gazit}}]{De-Leon:2016wyu}
\bibinfo{author}{\bibfnamefont{H.}~\bibnamefont{De-Leon}},
  \bibinfo{author}{\bibfnamefont{L.}~\bibnamefont{Platter}}, \bibnamefont{and}
  \bibinfo{author}{\bibfnamefont{D.}~\bibnamefont{Gazit}}
  (\bibinfo{year}{2016}), \eprint{1611.10004}.

\bibitem[{\citenamefont{Sadeghi et~al.}(2006)\citenamefont{Sadeghi, Bayegan,
  and Grie{\ss}hammer}}]{Sadeghi:2006fc}
\bibinfo{author}{\bibfnamefont{H.}~\bibnamefont{Sadeghi}},
  \bibinfo{author}{\bibfnamefont{S.}~\bibnamefont{Bayegan}}, \bibnamefont{and}
  \bibinfo{author}{\bibfnamefont{H.~W.} \bibnamefont{Grie{\ss}hammer}},
  \bibinfo{journal}{Phys. Lett.} \textbf{\bibinfo{volume}{B643}},
  \bibinfo{pages}{263} (\bibinfo{year}{2006}), \eprint{nucl-th/0610029}.

\bibitem[{\citenamefont{Arani et~al.}(2014)\citenamefont{Arani, Nematollahi,
  Mahboubi, and Bayegan}}]{Arani:2014qsa}
\bibinfo{author}{\bibfnamefont{M.~M.} \bibnamefont{Arani}},
  \bibinfo{author}{\bibfnamefont{H.}~\bibnamefont{Nematollahi}},
  \bibinfo{author}{\bibfnamefont{N.}~\bibnamefont{Mahboubi}}, \bibnamefont{and}
  \bibinfo{author}{\bibfnamefont{S.}~\bibnamefont{Bayegan}},
  \bibinfo{journal}{Phys. Rev.} \textbf{\bibinfo{volume}{C89}},
  \bibinfo{pages}{064005} (\bibinfo{year}{2014}), \eprint{1406.6530}.

\bibitem[{\citenamefont{Vanasse}(2017{\natexlab{a}})}]{Vanasse:2015fph}
\bibinfo{author}{\bibfnamefont{J.}~\bibnamefont{Vanasse}},
  \bibinfo{journal}{Phys. Rev.} \textbf{\bibinfo{volume}{C95}},
  \bibinfo{pages}{024002} (\bibinfo{year}{2017}{\natexlab{a}}),
  \eprint{1512.03805}.

\bibitem[{\citenamefont{Schiavilla et~al.}(1990)\citenamefont{Schiavilla,
  Pandharipande, and Riska}}]{Schiavilla:1990zz}
\bibinfo{author}{\bibfnamefont{R.}~\bibnamefont{Schiavilla}},
  \bibinfo{author}{\bibfnamefont{V.~R.} \bibnamefont{Pandharipande}},
  \bibnamefont{and} \bibinfo{author}{\bibfnamefont{D.~O.} \bibnamefont{Riska}},
  \bibinfo{journal}{Phys. Rev.} \textbf{\bibinfo{volume}{C41}},
  \bibinfo{pages}{309} (\bibinfo{year}{1990}).

\bibitem[{\citenamefont{Marcucci et~al.}(1998)\citenamefont{Marcucci, Riska,
  and Schiavilla}}]{Marcucci:1998tb}
\bibinfo{author}{\bibfnamefont{L.~E.} \bibnamefont{Marcucci}},
  \bibinfo{author}{\bibfnamefont{D.~O.} \bibnamefont{Riska}}, \bibnamefont{and}
  \bibinfo{author}{\bibfnamefont{R.}~\bibnamefont{Schiavilla}},
  \bibinfo{journal}{Phys. Rev.} \textbf{\bibinfo{volume}{C58}},
  \bibinfo{pages}{3069} (\bibinfo{year}{1998}), \eprint{nucl-th/9805048}.

\bibitem[{\citenamefont{Piarulli et~al.}(2013)\citenamefont{Piarulli, Girlanda,
  Marcucci, Pastore, Schiavilla, and Viviani}}]{Piarulli:2012bn}
\bibinfo{author}{\bibfnamefont{M.}~\bibnamefont{Piarulli}},
  \bibinfo{author}{\bibfnamefont{L.}~\bibnamefont{Girlanda}},
  \bibinfo{author}{\bibfnamefont{L.~E.} \bibnamefont{Marcucci}},
  \bibinfo{author}{\bibfnamefont{S.}~\bibnamefont{Pastore}},
  \bibinfo{author}{\bibfnamefont{R.}~\bibnamefont{Schiavilla}},
  \bibnamefont{and} \bibinfo{author}{\bibfnamefont{M.}~\bibnamefont{Viviani}},
  \bibinfo{journal}{Phys. Rev.} \textbf{\bibinfo{volume}{C87}},
  \bibinfo{pages}{014006} (\bibinfo{year}{2013}), \eprint{1212.1105}.

\bibitem[{\citenamefont{Vanasse and Phillips}(2017)}]{Vanasse:2016umz}
\bibinfo{author}{\bibfnamefont{J.}~\bibnamefont{Vanasse}} \bibnamefont{and}
  \bibinfo{author}{\bibfnamefont{D.~R.} \bibnamefont{Phillips}},
  \bibinfo{journal}{Few Body Syst.} \textbf{\bibinfo{volume}{58}},
  \bibinfo{pages}{26} (\bibinfo{year}{2017}), \eprint{1607.08585}.

\bibitem[{\citenamefont{Schmidt}(1937)}]{Schmidt1937}
\bibinfo{author}{\bibfnamefont{T.}~\bibnamefont{Schmidt}},
  \bibinfo{journal}{Zeitschrift f{\"u}r Physik} \textbf{\bibinfo{volume}{106}},
  \bibinfo{pages}{358} (\bibinfo{year}{1937}), ISSN \bibinfo{issn}{0044-3328},
  \urlprefix\url{http://dx.doi.org/10.1007/BF01338744}.

\bibitem[{\citenamefont{Phillips et~al.}(2000)\citenamefont{Phillips, Rupak,
  and Savage}}]{Phillips:1999hh}
\bibinfo{author}{\bibfnamefont{D.~R.} \bibnamefont{Phillips}},
  \bibinfo{author}{\bibfnamefont{G.}~\bibnamefont{Rupak}}, \bibnamefont{and}
  \bibinfo{author}{\bibfnamefont{M.~J.} \bibnamefont{Savage}},
  \bibinfo{journal}{Phys. Lett. B} \textbf{\bibinfo{volume}{473}},
  \bibinfo{pages}{209} (\bibinfo{year}{2000}), \eprint{nucl-th/9908054}.

\bibitem[{\citenamefont{de~Swart et~al.}(1995)\citenamefont{de~Swart,
  Terheggen, and Stoks}}]{deSwart:1995ui}
\bibinfo{author}{\bibfnamefont{J.~J.} \bibnamefont{de~Swart}},
  \bibinfo{author}{\bibfnamefont{C.~P.~F.} \bibnamefont{Terheggen}},
  \bibnamefont{and} \bibinfo{author}{\bibfnamefont{V.~G.~J.}
  \bibnamefont{Stoks}}, in \emph{\bibinfo{booktitle}{{3rd International
  Symposium on Dubna Deuteron 95 Dubna, Russia, July 4-7, 1995}}}
  (\bibinfo{year}{1995}), \eprint{nucl-th/9509032}.

\bibitem[{\citenamefont{Beane and Savage}(2001)}]{Beane:2000fi}
\bibinfo{author}{\bibfnamefont{S.~R.} \bibnamefont{Beane}} \bibnamefont{and}
  \bibinfo{author}{\bibfnamefont{M.~J.} \bibnamefont{Savage}},
  \bibinfo{journal}{Nucl. Phys.} \textbf{\bibinfo{volume}{A694}},
  \bibinfo{pages}{511} (\bibinfo{year}{2001}), \eprint{nucl-th/0011067}.

\bibitem[{\citenamefont{Kaplan et~al.}(1999)\citenamefont{Kaplan, Savage, and
  Wise}}]{Kaplan:1998sz}
\bibinfo{author}{\bibfnamefont{D.~B.} \bibnamefont{Kaplan}},
  \bibinfo{author}{\bibfnamefont{M.~J.} \bibnamefont{Savage}},
  \bibnamefont{and} \bibinfo{author}{\bibfnamefont{M.~B.} \bibnamefont{Wise}},
  \bibinfo{journal}{Phys. Rev.} \textbf{\bibinfo{volume}{C59}},
  \bibinfo{pages}{617} (\bibinfo{year}{1999}), \eprint{nucl-th/9804032}.

\bibitem[{\citenamefont{Wapstra and Audi}(1985)}]{Wapstra:1985zz}
\bibinfo{author}{\bibfnamefont{A.~H.} \bibnamefont{Wapstra}} \bibnamefont{and}
  \bibinfo{author}{\bibfnamefont{G.}~\bibnamefont{Audi}},
  \bibinfo{journal}{Nucl. Phys.} \textbf{\bibinfo{volume}{A432}},
  \bibinfo{pages}{1} (\bibinfo{year}{1985}).

\bibitem[{\citenamefont{Hagen et~al.}(2013)\citenamefont{Hagen, Hammer, and
  Platter}}]{Hagen:2013xga}
\bibinfo{author}{\bibfnamefont{P.}~\bibnamefont{Hagen}},
  \bibinfo{author}{\bibfnamefont{H.-W.} \bibnamefont{Hammer}},
  \bibnamefont{and} \bibinfo{author}{\bibfnamefont{L.}~\bibnamefont{Platter}},
  \bibinfo{journal}{Eur. Phys. J.} \textbf{\bibinfo{volume}{A49}},
  \bibinfo{pages}{118} (\bibinfo{year}{2013}), \eprint{1304.6516}.

\bibitem[{\citenamefont{Mehen et~al.}(1999)\citenamefont{Mehen, Stewart, and
  Wise}}]{Mehen:1999qs}
\bibinfo{author}{\bibfnamefont{T.}~\bibnamefont{Mehen}},
  \bibinfo{author}{\bibfnamefont{I.~W.} \bibnamefont{Stewart}},
  \bibnamefont{and} \bibinfo{author}{\bibfnamefont{M.~B.} \bibnamefont{Wise}},
  \bibinfo{journal}{Phys.Rev.Lett.} \textbf{\bibinfo{volume}{83}},
  \bibinfo{pages}{931} (\bibinfo{year}{1999}), \eprint{hep-ph/9902370}.

\bibitem[{\citenamefont{Schiff}(1964)}]{Schiff:1964zz}
\bibinfo{author}{\bibfnamefont{L.~I.} \bibnamefont{Schiff}},
  \bibinfo{journal}{Phys. Rev.} \textbf{\bibinfo{volume}{133}},
  \bibinfo{pages}{B802} (\bibinfo{year}{1964}).

\bibitem[{\citenamefont{Vanasse and Schindler}(2014)}]{Vanasse:2014sva}
\bibinfo{author}{\bibfnamefont{J.}~\bibnamefont{Vanasse}} \bibnamefont{and}
  \bibinfo{author}{\bibfnamefont{M.~R.} \bibnamefont{Schindler}},
  \bibinfo{journal}{Phys. Rev.} \textbf{\bibinfo{volume}{C90}},
  \bibinfo{pages}{044001} (\bibinfo{year}{2014}), \eprint{1404.0658}.

\bibitem[{\citenamefont{Cox et~al.}(1965)\citenamefont{Cox, Wynchank, and
  Collie}}]{Cox:1965}
\bibinfo{author}{\bibfnamefont{A.}~\bibnamefont{Cox}},
  \bibinfo{author}{\bibfnamefont{S.}~\bibnamefont{Wynchank}}, \bibnamefont{and}
  \bibinfo{author}{\bibfnamefont{C.}~\bibnamefont{Collie}},
  \bibinfo{journal}{Nuclear Physics} \textbf{\bibinfo{volume}{74}},
  \bibinfo{pages}{497 } (\bibinfo{year}{1965}), ISSN \bibinfo{issn}{0029-5582},
  \urlprefix\url{http://www.sciencedirect.com/science/article/pii/0029558265901975}.

\bibitem[{\citenamefont{Kirscher et~al.}(2010)\citenamefont{Kirscher,
  Grie{\ss}hammer, Shukla, and Hofmann}}]{Kirscher:2009aj}
\bibinfo{author}{\bibfnamefont{J.}~\bibnamefont{Kirscher}},
  \bibinfo{author}{\bibfnamefont{H.~W.} \bibnamefont{Grie{\ss}hammer}},
  \bibinfo{author}{\bibfnamefont{D.}~\bibnamefont{Shukla}}, \bibnamefont{and}
  \bibinfo{author}{\bibfnamefont{H.~M.} \bibnamefont{Hofmann}},
  \bibinfo{journal}{Eur. Phys. J.} \textbf{\bibinfo{volume}{A44}},
  \bibinfo{pages}{239} (\bibinfo{year}{2010}), \eprint{0903.5538}.

\bibitem[{\citenamefont{Patrignani et~al.}(2016)}]{Olive:2016xmw}
\bibinfo{author}{\bibfnamefont{C.}~\bibnamefont{Patrignani}}
  \bibnamefont{et~al.} (\bibinfo{collaboration}{Particle Data Group}),
  \bibinfo{journal}{Chin. Phys.} \textbf{\bibinfo{volume}{C40}},
  \bibinfo{pages}{100001} (\bibinfo{year}{2016}).

\bibitem[{\citenamefont{Lee et~al.}(2015)\citenamefont{Lee, Arrington, and
  Hill}}]{Lee:2015jqa}
\bibinfo{author}{\bibfnamefont{G.}~\bibnamefont{Lee}},
  \bibinfo{author}{\bibfnamefont{J.~R.} \bibnamefont{Arrington}},
  \bibnamefont{and} \bibinfo{author}{\bibfnamefont{R.~J.} \bibnamefont{Hill}},
  \bibinfo{journal}{Phys. Rev.} \textbf{\bibinfo{volume}{D92}},
  \bibinfo{pages}{013013} (\bibinfo{year}{2015}), \eprint{1505.01489}.

\bibitem[{\citenamefont{Epstein et~al.}(2014)\citenamefont{Epstein, Paz, and
  Roy}}]{Epstein:2014zua}
\bibinfo{author}{\bibfnamefont{Z.}~\bibnamefont{Epstein}},
  \bibinfo{author}{\bibfnamefont{G.}~\bibnamefont{Paz}}, \bibnamefont{and}
  \bibinfo{author}{\bibfnamefont{J.}~\bibnamefont{Roy}},
  \bibinfo{journal}{Phys. Rev.} \textbf{\bibinfo{volume}{D90}},
  \bibinfo{pages}{074027} (\bibinfo{year}{2014}), \eprint{1407.5683}.

\bibitem[{\citenamefont{Vanasse}(2017{\natexlab{b}})}]{Vanasse:2016hgn}
\bibinfo{author}{\bibfnamefont{J.}~\bibnamefont{Vanasse}},
  \bibinfo{journal}{Phys. Rev.} \textbf{\bibinfo{volume}{C95}},
  \bibinfo{pages}{024318} (\bibinfo{year}{2017}{\natexlab{b}}),
  \eprint{1609.08552}.

\bibitem[{\citenamefont{Amroun et~al.}(1994)}]{Amroun:1994qj}
\bibinfo{author}{\bibfnamefont{A.}~\bibnamefont{Amroun}} \bibnamefont{et~al.},
  \bibinfo{journal}{Nucl. Phys.} \textbf{\bibinfo{volume}{A579}},
  \bibinfo{pages}{596} (\bibinfo{year}{1994}).

\bibitem[{\citenamefont{Karshenboim}(2014)}]{Karshenboim:2014vea}
\bibinfo{author}{\bibfnamefont{S.~G.} \bibnamefont{Karshenboim}},
  \bibinfo{journal}{Phys. Rev.} \textbf{\bibinfo{volume}{D90}},
  \bibinfo{pages}{053013} (\bibinfo{year}{2014}), \eprint{1405.6515}.

\bibitem[{\citenamefont{Antognini}(2015)}]{Antognini:2015vxo}
\bibinfo{author}{\bibfnamefont{A.}~\bibnamefont{Antognini}}
  (\bibinfo{year}{2015}), \eprint{1512.01765}.

\bibitem[{\citenamefont{Franke et~al.}(2017)\citenamefont{Franke, Krauth,
  Antognini, Diepold, Kottmann, and Pohl}}]{Franke:2017tpc}
\bibinfo{author}{\bibfnamefont{B.}~\bibnamefont{Franke}},
  \bibinfo{author}{\bibfnamefont{J.~J.} \bibnamefont{Krauth}},
  \bibinfo{author}{\bibfnamefont{A.}~\bibnamefont{Antognini}},
  \bibinfo{author}{\bibfnamefont{M.}~\bibnamefont{Diepold}},
  \bibinfo{author}{\bibfnamefont{F.}~\bibnamefont{Kottmann}}, \bibnamefont{and}
  \bibinfo{author}{\bibfnamefont{R.}~\bibnamefont{Pohl}}
  (\bibinfo{year}{2017}), \eprint{1705.00352}.

\bibitem[{\citenamefont{Carlson et~al.}(2017)\citenamefont{Carlson, Gorchtein,
  and Vanderhaeghen}}]{Carlson:2016cii}
\bibinfo{author}{\bibfnamefont{C.~E.} \bibnamefont{Carlson}},
  \bibinfo{author}{\bibfnamefont{M.}~\bibnamefont{Gorchtein}},
  \bibnamefont{and}
  \bibinfo{author}{\bibfnamefont{M.}~\bibnamefont{Vanderhaeghen}},
  \bibinfo{journal}{Phys. Rev.} \textbf{\bibinfo{volume}{A95}},
  \bibinfo{pages}{012506} (\bibinfo{year}{2017}), \eprint{1611.06192}.

\bibitem[{\citenamefont{Nevo~Dinur et~al.}(2016)\citenamefont{Nevo~Dinur, Ji,
  Bacca, and Barnea}}]{Dinur:2015vzv}
\bibinfo{author}{\bibfnamefont{N.}~\bibnamefont{Nevo~Dinur}},
  \bibinfo{author}{\bibfnamefont{C.}~\bibnamefont{Ji}},
  \bibinfo{author}{\bibfnamefont{S.}~\bibnamefont{Bacca}}, \bibnamefont{and}
  \bibinfo{author}{\bibfnamefont{N.}~\bibnamefont{Barnea}},
  \bibinfo{journal}{Phys. Lett.} \textbf{\bibinfo{volume}{B755}},
  \bibinfo{pages}{380} (\bibinfo{year}{2016}), \eprint{1512.05773}.

\bibitem[{\citenamefont{Hernandez et~al.}(2016)\citenamefont{Hernandez,
  Nevo~Dinur, Ji, Bacca, and Barnea}}]{Hernandez:2016jlh}
\bibinfo{author}{\bibfnamefont{O.~J.} \bibnamefont{Hernandez}},
  \bibinfo{author}{\bibfnamefont{N.}~\bibnamefont{Nevo~Dinur}},
  \bibinfo{author}{\bibfnamefont{C.}~\bibnamefont{Ji}},
  \bibinfo{author}{\bibfnamefont{S.}~\bibnamefont{Bacca}}, \bibnamefont{and}
  \bibinfo{author}{\bibfnamefont{N.}~\bibnamefont{Barnea}},
  \bibinfo{journal}{Hyperfine Interact.} \textbf{\bibinfo{volume}{237}},
  \bibinfo{pages}{158} (\bibinfo{year}{2016}), \eprint{1604.06496}.

\bibitem[{\citenamefont{Pohl et~al.}(2013)\citenamefont{Pohl, Gilman, Miller,
  and Pachucki}}]{Pohl:2013yb}
\bibinfo{author}{\bibfnamefont{R.}~\bibnamefont{Pohl}},
  \bibinfo{author}{\bibfnamefont{R.}~\bibnamefont{Gilman}},
  \bibinfo{author}{\bibfnamefont{G.~A.} \bibnamefont{Miller}},
  \bibnamefont{and} \bibinfo{author}{\bibfnamefont{K.}~\bibnamefont{Pachucki}},
  \bibinfo{journal}{Ann. Rev. Nucl. Part. Sci.} \textbf{\bibinfo{volume}{63}},
  \bibinfo{pages}{175} (\bibinfo{year}{2013}), \eprint{1301.0905}.

\bibitem[{\citenamefont{Mohr et~al.}(2012)\citenamefont{Mohr, Taylor, and
  Newell}}]{Mohr:2012tt}
\bibinfo{author}{\bibfnamefont{P.~J.} \bibnamefont{Mohr}},
  \bibinfo{author}{\bibfnamefont{B.~N.} \bibnamefont{Taylor}},
  \bibnamefont{and} \bibinfo{author}{\bibfnamefont{D.~B.}
  \bibnamefont{Newell}}, \bibinfo{journal}{Rev. Mod. Phys.}
  \textbf{\bibinfo{volume}{84}}, \bibinfo{pages}{1527} (\bibinfo{year}{2012}),
  \eprint{1203.5425}.

\bibitem[{\citenamefont{Antognini et~al.}(2013)}]{Antognini:1900ns}
\bibinfo{author}{\bibfnamefont{A.}~\bibnamefont{Antognini}}
  \bibnamefont{et~al.}, \bibinfo{journal}{Science}
  \textbf{\bibinfo{volume}{339}}, \bibinfo{pages}{417} (\bibinfo{year}{2013}).

\end{thebibliography}

\end{document}